\documentclass{aa}  

\usepackage{graphicx}
\usepackage{txfonts}
\usepackage{lipsum}
\usepackage{color}
\usepackage[dvipsnames]{xcolor}
\usepackage{ulem}
\usepackage{siunitx}
\usepackage{ragged2e} 
\usepackage{multirow}
\usepackage{natbib}
\usepackage{bm}
\usepackage[colorlinks=true,citecolor=blue,linkcolor=red,filecolor=magenta,urlcolor=cyan]{hyperref}
\usepackage{subcaption}         
\usepackage{lscape}             
\usepackage{placeins}           
\makeatletter

\newcommand{\Rmnum}[1]{\expandafter\@slowromancap\romannumeral #1@}
\makeatother

\def\neworcid#1{\kern .08em\href{https://orcid.org/#1}{\includegraphics[keepaspectratio,width=0.7em]{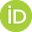}}}
             
\begin{document}

   \title{Evidence for stellar contamination and water absorption in NGTS-5b's transmission spectra with GTC/OSIRIS}

   \subtitle{}

   \titlerunning{Transmission spectra of NGTS-5b with GTC/OSIRIS}

   \author{Wan-Hao Wang\inst{\ref{pmo},\ref{ustc}}\neworcid{0009-0008-1371-229X} 
   \and Guo Chen\inst{\ref{pmo}}\neworcid{0000-0003-0740-5433} 
   \and Chengzi Jiang\inst{\ref{iac},\ref{ull}}\neworcid{0000-0003-1381-5527}
   \and Enric Pall\'e\inst{\ref{iac},\ref{ull}}\neworcid{0000-0003-0987-1593}
   \and Felipe Murgas\inst{\ref{iac},\ref{ull}}\neworcid{0000-0001-9087-1245}
   \and Hannu Parviainen\inst{\ref{ull},\ref{iac}}\neworcid{0000-0001-5519-1391}
          }

   \institute{
   CAS Key Laboratory of Planetary Sciences, Purple Mountain Observatory, Chinese Academy of Sciences, Nanjing, 210023, PR China\label{pmo}\\
   \email{guochen@pmo.ac.cn}
   \and School of Astronomy and Space Science, University of Science and Technology of China, Hefei, 230026, PR China\label{ustc}
   \and Instituto de Astrof\'isica de Canarias (IAC), V\'ia L\'actea s/n, 38205 La Laguna, Tenerife, Spain\label{iac}
   \and Departamento de Astrof\'isica, Universidad de La Laguna (ULL), C/ Padre Herrera, 38206 La Laguna, Tenerife, Spain\label{ull}}

   \date{Received November 10, 2025}

 
  \abstract
   {Transmission spectroscopy serves as a valuable tool for probing atmospheric absorption features in the terminator regions of exoplanets. Stellar surface heterogeneity can introduce wavelength-dependent contamination that complicates the interpretation of planetary spectra. }
   {We aim to investigate the atmosphere of the warm sub-Saturn NGTS-5b through optical transmission spectroscopy. }
   {Two transits were observed with the low-resolution Optical System for Imaging and low-Intermediate-Resolution Integrated Spectroscopy (OSIRIS) on the 10.4~m Gran Telescopio Canarias (GTC). Chromatic transit light curves were modeled to derive optical transmission spectra and multiple Bayesian spectral retrievals were performed to characterize the atmospheric properties.}
   {Model comparisons provide strong evidence for contamination from unocculted stellar spots. A joint retrieval of the transmission spectra, assuming equilibrium chemistry, indicates a relatively clear atmosphere with a sub-solar C/O ratio of $<$0.22 (90\% upper limit) and a low metallicity of $0.10^{+0.34}_{-0.05} \times$ solar. Retrievals assuming free chemistry yield strong evidence for the presence of $\rm H_2O$, with its abundance constrained to $\log X_{\mathrm{H_2O}} = -0.79^{+0.14}_{-0.17}$. However, the abundances of other species remain unconstrained due to the limited optical wavelength coverage. }
   {The discrepancies between the two NGTS-5b transit spectra can be attributed to varying levels of stellar contamination. NGTS-5b thus appears to host a relatively clear, water-rich atmosphere, pending confirmation from additional observations of molecular bands in the infrared.}


   \maketitle

\section{Introduction}
\label{Section:introduction}
Transmission spectroscopy \citep{Seager2000} has been the primary technique for probing absorption and scattering features in exoplanet atmospheres, particularly at the planet's day-night terminator. To date, precise spectroscopic measurements have been obtained for over 200 exoplanets\footnote{\url{http://research.iac.es/proyecto/exoatmospheres/index.php}}, with a primary focus on close-in giant worlds. Water has been detected in nearly half of these atmospheres \citep{Madhusudhan2019}, making it a key tracer for atmospheric composition and chemical processes. 

The launch of the James Webb Space Telescope (JWST) in 2021 has enabled mid-infrared observations with unprecedented sensitivity, allowing the identification of a range of key molecules in exoplanet atmospheres, such as $\rm H_2O$, $\rm CH_4$, $\rm CO_2$, $\rm SO_2$, and $\rm H_2S$ \citep[e.g.,][]{Rustamkulov2023,Bell2023,Dyrek2024,Powell2024,Kirk2025}. However, JWST offers limited coverage at blue-optical wavelengths, where scattering slopes and pressure-broadened sodium lines are most prominent \citep{Nikolov2018,Radica2023}. Complementary observations from large ground-based telescopes are therefore essential to provide crucial constraints on stellar contamination and atmospheric properties. 

Interpreting transmission spectra requires a thorough understanding of the wavelength-dependent brightness of the host star. However, stellar magnetic activity, in the form of spots (cooler, darker regions; for details, see review by \citealp{Solanki2003}) and faculae (hotter, brighter regions; see review by \citealp{Unruh1999}), can produce surface heterogeneity across the stellar disk. When these active regions are occulted by the transiting planet, they can imprint detectable anomalies in the light curve \citep[e.g.,][]{Barros2013, Mohler-Fischer2013, Kirk2016, Baluev2021, Libby-Roberts2023}. In addition, unocculted active regions can bias transmission spectra by altering the baseline stellar flux \citep[e.g.,][]{Rackham2017,Jiang2021,Lim2023,Moran2023,Banerjee2024}, which commonly referred to as the transit light source (TLS) effect \citep{Apai2018}. 

NGTS-5b is a warm sub-Saturn planet with a mass of $\sim 0.229$~$M_{\rm Jup}$, a radius of $\sim 1.136$~$R_{\rm Jup}$, and an equilibrium temperature of $\sim 952$~K \citep{Eigmuller2019}. It orbits its host star with a period of $\sim 3.357$~days, placing it at the upper boundary of the sub-Jovian desert \citep{Szabo2011}. The host star is a K2-type dwarf with a mass of $\sim 0.661$~$M_\odot$, a radius of $\sim 0.739$~$R_\odot$, an effective temperature of $\sim 4987$~K, and a metallicity of $\sim 0.12$~dex. Notably, a photometric survey targeting the metastable helium triplet at 10830~$\AA$ with Palomar/WIRC tentatively detected absorption signals from NGTS-5b, highlighting it as a promising candidate for investigating atmospheric escape processes and for assessing the influence of stellar activities \citep{Vissapragada2022}.

In this study, we present the first optical transmission spectroscopy of NGTS-5b, based on two transit observations obtained with the Optical System for Imaging and low-Intermediate-Resolution Integrated Spectroscopy (OSIRIS; \citealt{Cepa2000}) on the 10.4~m Gran Telescopio CANARIAS (GTC). Our primary aim is to characterize the planet's atmospheric properties by identifying absorption and scattering features in its optical transmission spectrum. 

This paper is structured as follows. Section \ref{Section:observation} summarizes the observations and data reduction procedures. Section \ref{Section:lc} presents the analysis of the transit light curves. In Section \ref{Section:retrieval}, we perform atmospheric retrievals and interpret the resulting transmission spectra. Section \ref{Section:discussion} discusses possible alternative interpretations and compares our findings with those for similar exoplanets. Finally, Section \ref{Section:conclusions} summarizes our conclusions.

\section{Observations and data reduction}
\label{Section:observation}

\begin{table}
\centering
\renewcommand\arraystretch{1.5}
\caption{Observation summary.} 
\label{tab:obs_sum}
\scalebox{0.85}{
\begin{tabular}{lcc}
\hline\noalign{\smallskip}
Parameter & Night 1 & Night 2 \\\noalign{\smallskip}
\hline\noalign{\smallskip}
Program ID & GTC24-20A & GTC1-19ITP \\
PI & \multicolumn{2}{c}{E. Pallé} \\
RA (J2000) & \multicolumn{2}{c}{$\mathrm{14^h 44^m 13.98^s}$} \\
DEC (J2000) & \multicolumn{2}{c}{$+05^\circ 36^{\prime} 19.3^{\prime\prime}$} \\
Observing date (UT) & 2020-03-10 & 2021-03-21 \\
Start time (UT) & 01:20 & 00:23 \\
End time (UT) & 06:00 & 06:24 \\
Exposure time (s) & 65 & 55 \\
Useful exposures & 174 & 269 \\
Airmass \tablefootmark{a} & 1.669 -- 1.088 -- 1.147 & 1.812 -- 1.088 -- 1.326 \\
Seeing \tablefootmark{b} & $1.15^{\prime\prime}$ -- $3.25^{\prime\prime}$ & $0.86^{\prime\prime}$ -- $2.02^{\prime\prime}$ \\\noalign{\smallskip}
\hline
\end{tabular}}
\tablefoot{
\tablefoottext{a}{Start – minimum – end}
\tablefoottext{b}{95\% interval, measured by the full width at half maximum (FWHM) of the stars' point spread function (PSF) in the spatial direction at the central wavelengths of 7195\si{\angstrom} for Night 1 and 5765\si{\angstrom} for Night 2. }
}
\end{table}

\subsection{GTC/OSIRIS transit spectroscopy}
We observed two transits of NGTS-5b using GTC OSIRIS on the nights of 10 March 2020 (Night 1) and 21 March 2021 (Night 2). OSIRIS is equipped with two Marconi CCD44-82 detectors ($2048\times4096$~pixels each), providing an unvignetted field of view (FOV) of $\sim$$7.8^{\prime}$. 

For both nights, the target star (NGTS-5, 2MASS 14441396+0536195) was observed simultaneously with a reference star (2MASS 14440727+0530541) using a $7.4^{\prime}$-long and $40^{\prime\prime}$-wide slit. The target and reference stars have $r$-band magnitudes of 13.483 and 12.965, respectively, with an angular separation of $5.67^{\prime}$ \citep{Zacharias2012}. The target was placed on CCD 1 and the reference star on CCD 2 during both observations. The CCDs were read out in the standard 200~kHz mode with $2 \times 2$ binning, resulting in a spatial scale of $0.254^{\prime\prime}$ per binned pixel. The R1000R grism was employed on Night 1, covering a wavelength range from 513 to 1043~nm with a dispersion of about 2.62~$\AA$ per binned pixel. On Night 2, the R1000B grism was used, spanning 364 to 789~nm with a dispersion of about 2.12~$\AA$ per binned pixel. 

The first transit was observed from UT 01:20 to 06:00 on Night 1, with seeing ranging from $1.15^{\prime\prime}$ to $3.25^{\prime\prime}$ and a median value of $1.66^{\prime\prime}$. The second transit was observed from UT 00:23 to 06:24 on Night 2, with seeing varying between $0.86^{\prime\prime}$ and $2.02^{\prime\prime}$ and a median value of $1.29^{\prime\prime}$. The first 13 exposures of Night 1 were excluded from further analyses due to a systematic wavelength shift, which was likely caused by a displacement of the stars across the slit width. 
The last seven exposures of Night 2 were discarded due to an anomalous flux drop, which was possibly caused by a partial obstruction in the telescope’s optical path. A summary of the observational setup and conditions is provided in Table \ref{tab:obs_sum}. The data are now publicly available through the GTC Archive\footnote{\url{https://gtc.sdc.cab.inta-csic.es/gtc/index.jsp}}.

\subsection{Data reduction}
\begin{figure}
\centering
\includegraphics[width=1.0\hsize]{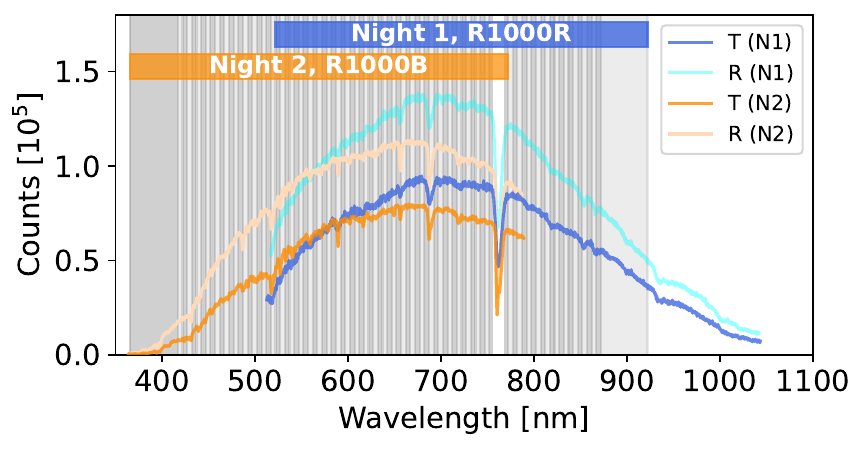}
\caption{Example stellar spectra of NGTS-5 (T; darker line) and its reference star (R; lighter line), observed with the R1000R grism on Night 1 (blue) and the R1000B grism on Night 2 (orange). The vertical lines with gray-shaded regions indicate the adopted spectroscopic passbands. }
\label{fig:t&r_flux}
\end{figure}

The reduction of raw spectral images obtained with OSIRIS followed the established procedures outlined in our previous works \citep{Chen2017b,Chen2017a,Chen2018,Chen2020a,Chen2021a,Parviainen2018,Murgas2019,Murgas2020,Jiang2021,Jiang2023,Jiang2024,Kang2024}. The two-dimensional spectral images were processed through overscan correction, bias subtraction, flat-field calibration, cosmic-ray removal, and sky background subtraction. One-dimensional spectra were extracted using the optimal extraction algorithm \citep{Horne1986}, implemented via the \texttt{IRAF APALL} package \citep{Tody1993}. The optimal aperture radius was determined by minimizing the scatter in the white-light curves, resulting in 12.0~pixels for Night 1 and 11.5~pixels for Night 2. Preliminary wavelength calibration was performed using HeAr, Ne, and Xe arc lamps with R1000R for Night 1, and HeAr and Ne arc lamps with R1000B for Night 2. Both calibrations employed a $1.23^{\prime\prime}$-wide slit. 

Subsequently, we used customized \texttt{IDL} routines to generate spectral cubes for both the target and reference stars. Wavelength solutions were refined by correcting shifts in the $\mathrm{H\alpha}$ and Na \Rmnum{1} D lines relative to their laboratory air wavelengths for each exposure. The reference star spectra were further aligned to those of the target star by cross-correlating telluric absorption features. All time stamps ($t$) were converted to Barycentric Julian Date in the Barycentric Dynamical standard ($\rm BJD_{TDB}$; \citealp{Eastman2010}). Additionally, we tracked the spectra’s spatial and spectral drifts ($x$ and $y$), FWHM of the stellar profile ($s_y$), and the instrumental rotation angle ($\theta$), which were later incorporated into the light-curve modeling. 

For Night 1, the white light curve was constructed by summing the flux between 522–922~nm, excluding the 757-767~nm region to avoid contamination from the telluric oxygen A-band. The target star’s light curve was divided by that of the reference star to correct for telluric variations and then normalized to its out-of-transit flux level. Spectroscopic light curves were generated by binning the spectra into 70 narrow channels, typically 5~nm wide, with one broader 50~nm bin at the red end where the flux is lower. Wavelengths beyond 922~nm were discarded to avoid fringing effects. 

Similarly, for Night 2, the white light curve was created covering 366-772~nm, again excluding the telluric oxygen A-band. The spectroscopic light curves were divided into 71 channels, most of which are 5~nm wide, except for a 51~nm channel at the blue end. An example of the extracted 1D stellar spectra for both the target and reference stars, along with the adopted spectroscopic passbands, is shown in Figure \ref{fig:t&r_flux}. 

\begin{table}
\renewcommand\arraystretch{1.5}
\centering
\caption{Priors and derived posteriors for the parameters used in the white light curve analyses. }
\label{tab:wlc}
\scalebox{0.85}{
\begin{tabular}{ccc}
\hline
Parameter & Prior & Posterior Estimate \\
\hline
$P$ [d] & fixed & 3.3569866 \\
$i$ [$^{\circ}$] & $\mathcal{U}(80, 90)$ & $87.377^{+0.091}_{-0.088}$\\
$a/R_{\star}$ & $\mathcal{U}(5, 20)$ & $12.262^{+0.145}_{-0.130}$\\
$R_{\rm p}/R_{\star}$ & $\mathcal{U}(0, 0.3)$ & $0.14861^{+0.00191}_{-0.00204}$\\
\noalign{\medskip}
\hline
\noalign{\smallskip}
\multicolumn{3}{c}{2020-03-10 (Night 1)} \\
\noalign{\smallskip}
$t_c$ [MJD\tablefootmark{a}] & $\mathcal{U}(-0.02,0.02)$ & $0.00109^{+0.00040}_{-0.00038}$\\
$u_1$ & fixed & 1.06645 \\
$u_2$ & fixed & $-1.59440$ \\
$u_3$ & fixed & 2.28615 \\
$u_4$ & fixed & $-0.89105$ \\
$c_0$ & $\mathcal{U}(-\infty, +\infty)$ & $1.00852^{+0.00044}_{-0.00045}$\\
$c_1$ & $\mathcal{U}(-\infty, +\infty)$ & $-0.08214^{+0.00300}_{-0.00292}$\\
$c_2$ & $\mathcal{U}(-\infty, +\infty)$ & $-0.41182^{+0.08536}_{-0.08908}$\\
$\sigma_w$~$[10^{-6}]$ & $\mathcal{U}(0.1, 5000)$ & $0.00046^{+0.00004}_{-0.00003}$\\
$\ln \sigma_k$ & $\mathcal{U}(-10,-1)$ & $-7.06447^{+0.27910}_{-0.37981}$\\
$\ln \tau_t$ & $\mathcal{U}(-6,5)$ & $-3.79902^{+0.34960}_{-0.37174}$\\
$\ln \tau_x$ & $\mathcal{U}(-5,5)$ & $3.08645^{+1.05288}_{-1.34064}$\\
$\ln \tau_y$ & $\mathcal{U}(-5,5)$ & $3.03125^{+1.22712}_{-1.24824}$\\
$\ln \tau_{s_y}$ & $\mathcal{U}(-5,5)$ & $1.87409^{+0.66805}_{-0.45994}$\\
$\ln \tau_\theta$ & $\mathcal{U}(-5,5)$ & $3.10147^{+0.94035}_{-0.92081}$\\
\noalign{\medskip}
\hline
\noalign{\smallskip}
\multicolumn{3}{c}{2021-03-21 (Night 2)} \\
\noalign{\smallskip}
$t_c$ (MJD\tablefootmark{a}) & $\mathcal{U}(-0.02,0.02)$ & $0.00179^{+0.00015}_{-0.00015}$ \\
$u_1$ & fixed & 1.06645 \\
$u_2$ & fixed & $-1.59440$ \\
$u_3$ & fixed & 2.28615 \\
$u_4$ & fixed & $-0.89105$ \\
$c_0$ & $\mathcal{U}(-\infty, +\infty)$ & $0.99822^{+0.00017}_{-0.00018}$\\
$c_1$ & $\mathcal{U}(-\infty, +\infty)$ & $0.03096^{+0.00090}_{-0.00095}$\\
$c_2$ & $\mathcal{U}(-\infty, +\infty)$ & $0.18407^{+0.02176}_{-0.02165}$\\
$\sigma_w$~$[10^{-6}]$ & $\mathcal{U}(0.1, 5000)$ & $0.00034^{+0.00002}_{-0.00002}$\\
$\ln \sigma_k$ & $\mathcal{U}(-10,-1)$ & $-6.32446^{+0.53665}_{-0.84631}$\\
$\ln \tau_t$ & $\mathcal{U}(-6,5)$ & $-1.21432^{+0.85693}_{-1.10322}$\\
$\ln \tau_x$ & $\mathcal{U}(-5,5)$ & $4.15940^{+0.57947}_{-1.22978}$\\
$\ln \tau_y$ & $\mathcal{U}(-5,5)$ & $3.55239^{+1.00672}_{-1.41149}$\\
$\ln \tau_{s_y}$ & $\mathcal{U}(-5,5)$ & $2.18247^{+0.53245}_{-0.85872}$\\
$\ln \tau_\theta$ & $\mathcal{U}(-5,5)$ & $4.14256^{+0.42534}_{-0.45817}$\\
\noalign{\medskip}
\hline
\end{tabular}
}
\tablefoot{
\tablefoottext{a}{$\rm{MJD=BJD_{TDB} - 2457740.35262}.$}
}
\end{table}

\begin{figure*}
\centering
\includegraphics[width=1.0\hsize]{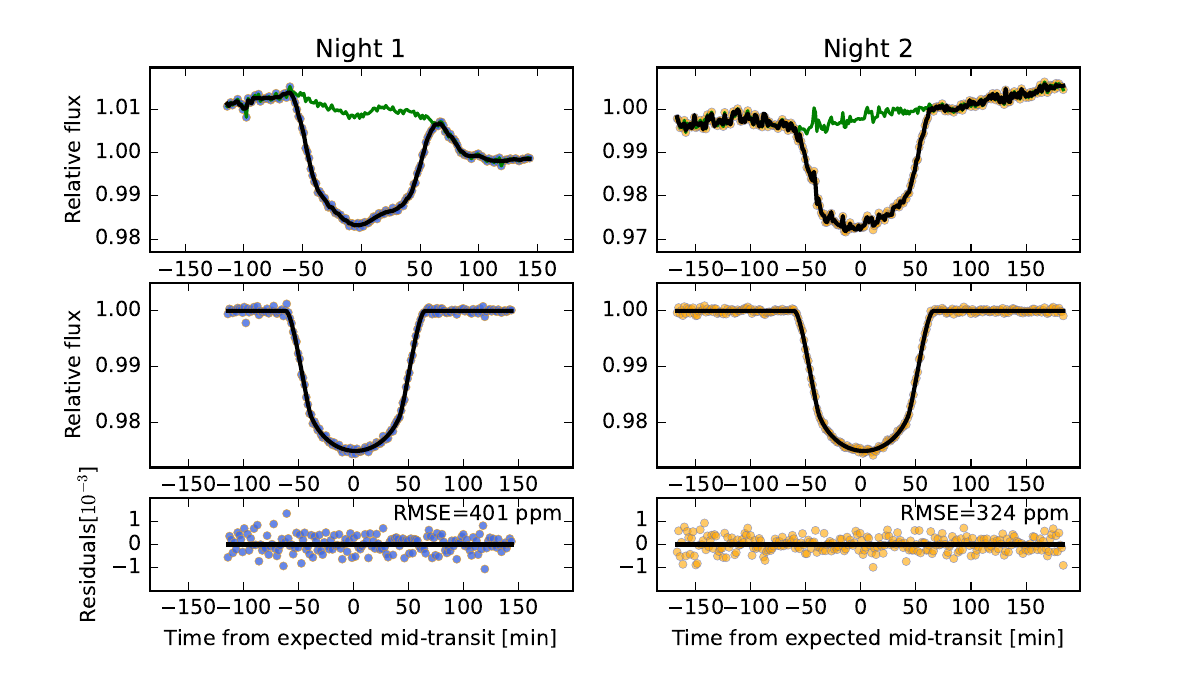}
\caption{White light curves of NGTS-5 observed with GTC/OSIRIS on Night 1 (left) and Night 2 (right). Top panels: observed white light curves (blue circles for Night 1, orange circles for Night 2), best-fit models (black lines), and modeled systematics (green lines). Middle panels: detrended white light curves with the best-fit transit models. Bottom panels: residuals between the observed data and best-fit models.  }
\label{fig:wlc}
\end{figure*}

\section{Light curve analysis}
\label{Section:lc}
The white and spectroscopic light curves were modeled following the approach of \citet{Chen2018}. Each light curve was assumed to consist of a transit signal superimposed with correlated systematics. 

To model the transit signal, we employed the Python package \texttt{batman} \citep{Kreidberg2015}, assuming a circular orbit. The transit model was parameterized by the orbital period ($P$), orbital inclination ($i$), scaled semimajor axis ($a/R_{\star}$), planet-to-star radius ratio ($R_{\rm p}/R_{\star}$), mid-transit time ($t_{\rm c}$), and stellar limb-darkening coefficients (LDCs). A four-parameter nonlinear limb-darkening law was adopted, with fixed coefficients ($u_1$, $u_2$, $u_3$, $u_4$) derived from the PHOENIX stellar atmosphere models \citep{Husser2013} via the Python tool of \citet{Espinoza2015}. Light curves were then computed using the efficient numerical integration scheme implemented in \texttt{batman} for higher-order limb-darkening models. The stellar model grid was configured to match NGTS-5’s effective temperature ($T_{\rm eff}=5000$~K), surface gravity ($\log g_{\star}=4.5$), and metallicity (${\rm [Fe/H]}=0.0$).

The correlated systematics were modeled using Gaussian processes (GP; \citealp{Rasmussen2006, Gibson2012}), implemented via the Python package \texttt{george} \citep{Ambikasaran2015}. The GP mean function was defined as the product of the transit model and a polynomial trend in time, i.e., $B = \sum_{i=0}^N c_it^{i}$. The GP covariance function was specified by a $3/2$-order Matérn kernel, incorporating multiple state vectors as input variables:
\begin{equation}
 k_{\rm total}(\mathbf{x}_i, \mathbf{x}_j) = \sigma_k^2 \prod_{\alpha} k_{\rm M32} r_{\alpha}\;,
 \label{eq:gpkernel}
\end{equation}
where $\sigma_k$ is the covariance amplitude, $\alpha$ denotes the index of individual GP input variables, $r_{\alpha} = |x_{\alpha,i} - x_{\alpha,j}|/\tau_{\alpha}$ represents the distance between the $i$-th and $j$-th inputs along the $\alpha$-th dimension, normalized by its characteristic length scale $\tau_{\alpha}$ for each input variable. The one-dimensional $3/2$-order Matérn kernel is given by: 
\begin{equation}
 k_{\rm M32}(r) = \left(1 + \sqrt{3} r \right) \exp\left( - \sqrt{3} r \right).
 \label{eq:M32}
\end{equation}
Prior to modeling, each state vector was standardized to zero mean and unit variance. To account for potential underestimation of photometric uncertainties, an additional white noise variance term $\sigma_w^2$ was added to the diagonal of the covariance matrix. The GP regression thus included the free parameters $\sigma_k$, $\tau_\alpha$, and $\sigma_w$. 

Finally, the posterior distributions of the model parameters were explored using the affine-invariant Markov Chain Monte Carlo (MCMC) algorithm, implemented with the Python package \texttt{emcee} \citep{Foreman-Mackey2013}.

\subsection{White light curves}
\label{Section:wlc}
The white light curves from both nights were jointly fitted, adopting common transit parameters except for the mid-transit time $t_{\rm c}$, while allowing separate quadratic trend and GP parameters to account for night-specific systematics. For both transits, all five state vectors \{$t$, $x$, $y$, $s_y$, $\theta$\} were selected as input variables for the GP covariance matrix. This resulted in a total of 25 free parameters in the joint light curve fitting. 

The orbital period was fixed to $P = 3.3569866$~days, as reported by \citet{Eigmuller2019}, and the LDCs were fixed to their precomputed values. Uniform priors were adopted for the remaining transit parameters \{$i$, $a/R_{\star}$, $R_{\rm p}/R_{\star}$, $t_{\rm c}$\}, trend coefficients \{$c_0$, $c_1$, $c_2$\}, and $\sigma_w$, while log-uniform priors were assigned to the other GP hyperparameters \{$\sigma_k$, $\tau_t$, $\tau_x$, $\tau_y$, $\tau_{s_y}$, $\tau_\theta$\}. Three chains with 60 walkers each were initialized in the MCMC process, with two short burn-in phases of 1,000-step, followed by a production run of 5,000 steps. Convergence was verified by ensuring the chain length significantly exceeded 50 times the integrated autocorrelation time. 

Figure \ref{fig:wlc} shows the observed white light curves, along with the best-fit transit models and systematics. The root mean square error (RMSE) of the residuals is 386~ppm for Night 1 and 321~ppm for Night 2, corresponding to 2.8 and 2.1 times the photon noise limit, respectively. The adopted priors and resulting posteriors for the free parameters are summarized in Table \ref{tab:wlc}. Our joint fit yield transit parameters of $i = 87.377^{+0.091}_{-0.088}\,^{\circ}$ and $a/R_{\star} = 12.262^{+0.145}_{-0.130}$, which deviate from the values reported in the discovery paper ($i = 86.6 \pm 0.2^{\circ}$, $a/R_{\star} = 11.111^{+0.315}_{-0.296}$; \citealp{Eigmuller2019}) at the $3.6\sigma$ and $3.4\sigma$ levels, respectively. This discrepancy is likely attributable to the well-known degeneracy between $i$ and $a/R_{\star}$ in transit light curve modeling, where a higher inclination tends to correlate positively with a larger scaled semimajor axis. 

\begin{figure*}
\centering
\includegraphics[width=0.49\hsize]{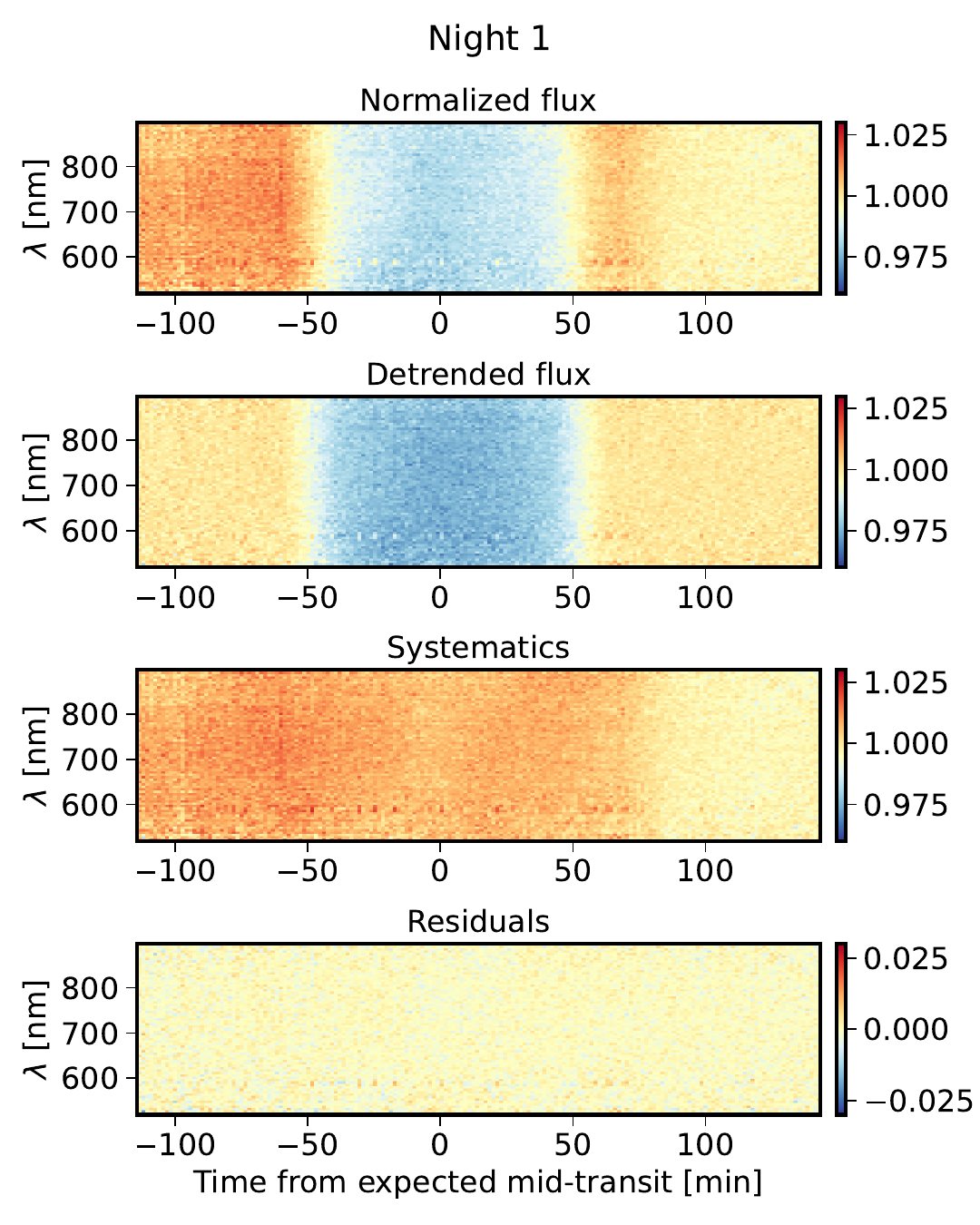}
\includegraphics[width=0.49\hsize]{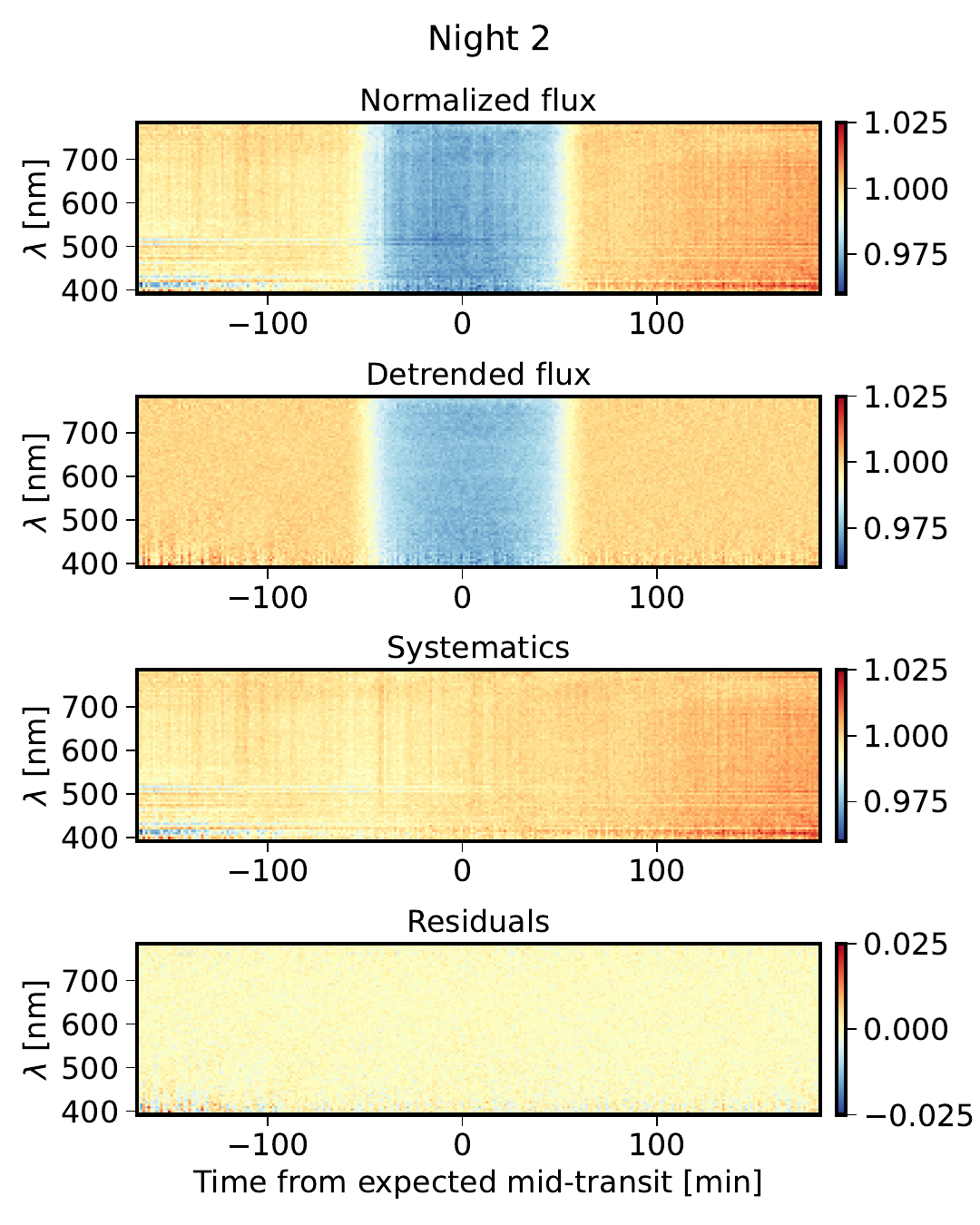}
\caption{Spectroscopic light curves of NGTS-5 observed with GTC/OSIRIS on Night 1 (left) and Night 2 (right). First row: matrix of the raw light curves. Second row: matrix of the detrended light curves. Third row: matrix of the extracted systematics. Fourth row: matrix of the residuals.}
\label{fig:slc_2D}
\end{figure*}

\begin{figure}
\centering
\includegraphics[width=1.0\hsize]{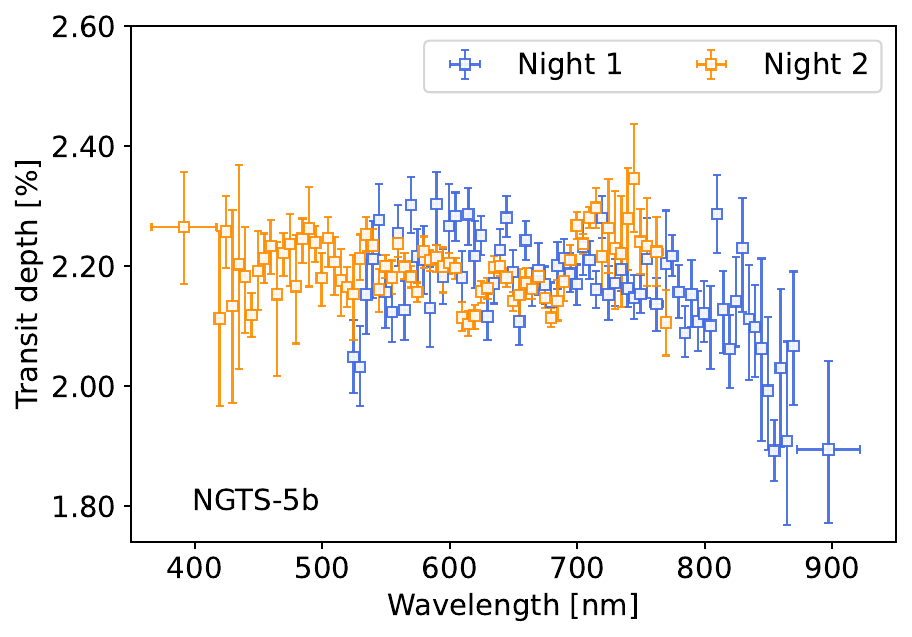}
\caption{Individual transmission spectra of NGTS-5b derived from Night 1 (blue) and Night 2 (orange) observations.}
\label{fig:slc}
\end{figure}

\subsection{Spectroscopic light curves}
\label{Section:slc}
Prior to modeling the spectroscopic light curves, we removed wavelength-independent systematics, known as common-mode systematics, to reduce the overall noise level \citep{Gibson2013b,Gibson2013a,Gibson2017,Stevenson2014,Chen2017b,Chen2017a}. For each transit, the common-mode systematics were derived by dividing the raw white light curve by its best-fit transit model. The resulting common-mode trend was then used to normalize each individual spectroscopic light curve for both nights. 

After correcting for common-mode systematics, the spectroscopic light curves were individually fitted for each night. For a given transit, each spectroscopic channel was modeled independently, with wavelength-dependent transit parameters and systematics parameters. The LDCs were fixed to the precomputed values. The other transit parameters \{$i$, $a/R_{\star}$, $t_{\rm c}$\} were fixed to the median values derived from the joint white light curves analysis (Table \ref{tab:wlc}). Two coefficients \{$c_0$, $c_1$\} were used to describe a linear time trend in the GP mean function, and three state vectors \{$t$, $y$, $s_y$\} were selected as inputs to the GP covariance matrix, resulting in eight free parameters for each spectroscopic light curve. Three chains with 32 walkers each were initialized in the MCMC process, with two short burn-in phases of 1,000 step, followed by a production run of 5,000 steps. 

Figure \ref{fig:slc_2D} shows the matrix of spectroscopic light curves along with the best-fit light curve systematics models and the corresponding residuals. The derived transmission spectra, along with the fixed LDCs for each spectroscopic passband, are presented in Table \ref{tab:spectra}. Figure \ref{fig:slc} shows the individual transmission spectra for the two nights. The two spectra exhibit noticeable differences, with a chi-squared value of $\chi^2 = 103.03$ for dof = 49 in the common passbands. In the next section, we investigate the possible role of stellar activity in explaining these differences.

\section{Atmospheric retrieval analysis}
\label{Section:retrieval}
\subsection{Model setup}
\subsubsection{Atmospheric forward model}
To characterize the atmospheric properties of NGTS-5b, we performed a series of Bayesian spectral retrieval analyses on its transmission spectra. The 1D forward models were generated using the Python package \texttt{petitRADTRANS 2.6.7} \citep{Molliere2019}. Bayesian parameter estimation and model comparison were conducted with the multimodal nested sampling algorithm implemented in \texttt{PyMultiNest} \citep{Buchner2014}, providing posterior distributions of model parameters and Bayesian evidences ($\mathcal{Z}$). 

The atmospheric temperature profile was assumed to be isothermal, with a temperature of $T_{\rm iso}$. The pressure range from $10^2$ to $10^{-6}$~${\rm bar}$ was equally divided into 100 layers on a logarithmic scale. The reference pressure $P_0$, corresponding to the planetary radius ($R_{\rm p} = 1.136$~$R_{\rm Jup}$), was treated as a free parameter to anchor the vertical structure. 

Key opacity sources included gas absorption, collision-induced absorption (CIA), Rayleigh scattering, and parameterized cloud and haze contributions. The atmospheric bulk composition was assumed to be dominated by $\rm H_2$ and $\rm He$ with a $3:1$ mass ratio, consistent with a roughly primordial composition \citep{Molliere2019}. Opacity contributions from 15 molecular and atomic species ($\rm Na$, $\rm K$, $\rm TiO$, $\rm VO$, $\rm H_2O$, $\rm CH_4$, $\rm CO$, $\rm CO_2$, $\rm HCN$, $\rm C_2H_2$, $\rm PH_3$, $\rm NH_3$, $\rm H_2S$, $\rm SiO$, and $\rm FeH$) were included. Continuum opacity sources included CIA from $\rm H_2 - H_2$ and $\rm H_2 - He$ pairs, and Rayleigh scattering by $\rm H_2$ and $\rm He$. 

Two different approaches were employed for handling gas abundances:
\begin{itemize}
    \item Equilibrium chemistry: where the mass fractions were interpolated from precomputed chemical grids \citep{Molliere2019}, parameterized by the atmospheric metallicity (${\rm log}Z$) and carbon-to-oxygen number ratio ($\rm C/O$).
    \item Free chemistry: where the mass fractions ($X_i$) of the 15 individual opacity species were treated as independent free parameters. 
\end{itemize}
Following the prescription of \citet{MacDonald2017}, clouds were modeled as a fractional coverage ($\phi$) of the terminator region. The cloudy region was modeled as an opaque cloud deck parameterized by its cloud-top pressure ($P_{\rm cloud}$), with an additional Rayleigh-like scattering component that enhances the short-wavelength opacity, characterized by an amplitude factor ($A_{\rm RS}$). Additionally, a free vertical offset ($\Delta_{\rm offset}$) was introduced between the two transits to correct for potential systematic differences. 

\subsubsection{Stellar contamination model}
We further accounted for the potential contamination from stellar surface heterogeneities, specifically spots and faculae, in the retrieval analyses. These introduce wavelength-dependent variations in the observed transmission spectra due to the temperature contrast between regions in and out of the transit chord. The transit chord was simply assumed quiescent, as potential signatures of spot or facula crossings could not be distinguished given the level of systematics in the light curves. A wavelength-dependent correction factor, $\beta_{\lambda}$, was introduced: 
\begin{equation}
 D_{\lambda, \rm c} = D_{\lambda}\beta_{\lambda}\;,
 \label{eq:TDc}
\end{equation}
where $D_{\lambda}$ and $D_{\lambda, \rm c}$ are the modeled transit depth without and with stellar contamination, respectively. 

The correction factor $\beta_{\lambda}$ depends on the relative flux contributions from the photosphere, spots, and faculae, with temperatures of $T_{\rm phot}$, $T_{\rm spot}$, and $T_{\rm facu}$ at coverage fractions of $1 - f_{\rm spot} - f_{\rm facu}$, $f_{\rm spot}$ and $f_{\rm facu}$: 
\begin{equation}
 \begin{split}
 \beta_{\lambda} &= S(\lambda, T_{\rm phot}) \; / \; \left[ f_{\rm spot}S(\lambda, T_{\rm spot}) \right. \\ &\left. + f_{\rm facu}S(\lambda, T_{\rm facu}) + (1 - f_{\rm spot} - f_{\rm facu})S(\lambda, T_{\rm phot}) \right]\;,
 \end{split}
 \label{eq:spot&facu}
\end{equation}
where $S(\lambda, T)$ denotes the stellar spectrum interpolated from the PHOENIX spectral library \citep{Husser2013}. 

To comprehensively assess the scenarios of planetary atmosphere and stellar contamination, we performed retrieval analyses under the following model assumptions: 
\begin{itemize}
    \item A. Null model: no atmosphere and no stellar contamination, i.e., a flat-line model.
    \item B. Pure atmospheric model with equilibrium chemistry. 
    \item C. Pure atmospheric model with free chemistry. 
    \item D. Pure stellar contamination model.
    \item E. Hybrid model: equilibrium chemistry atmosphere with stellar contamination. 
    \item F. Hybrid model: free chemistry atmosphere with stellar contamination. 
\end{itemize}

For model assumptions E and F, atmospheric properties remained the same for both transits while stellar contamination parameters ($\Delta T_{\rm spot} = T_{\rm spot} - T_{\rm phot}$, $\Delta T_{\rm facu} = T_{\rm facu} - T_{\rm phot}$, $f_{\rm spot}$, and $f_{\rm facu}$) were retrieved independently for each transit, considering the \textasciitilde 1-year gap between the two observations. 

Each retrieval run used 250 live points in \texttt{PyMultiNest}, typically yielding an uncertainty of $\sim$$0.2$ in $\ln\mathcal{Z}$. For model comparison, Bayes factor $\mathcal{B}_{10}$ ($\ln\mathcal{B}_{10} = \ln\mathcal{Z}_1 - \ln\mathcal{Z}_0 = \Delta\ln\mathcal{Z}_{10}$) was interpreted following the criteria of \citet{Trotta2008}: strong evidence for model 1 over model 0 if $\Delta\ln\mathcal{Z}_{10} \geq 5.0$, moderate if $2.5 \leq \Delta\ln\mathcal{Z}_{10} < 5.0$, weak if $1.0 \leq \Delta\ln\mathcal{Z}_{10} < 2.5$, and inconclusive if $\Delta\ln\mathcal{Z}_{10} < 1.0$. 

\begin{table}
\centering
\caption{Summary of Bayesian spectral retrieval statistics. }
\label{tab:model_cmp}
\scalebox{0.85}{
\begin{tabular}{ccccccc}
\hline\noalign{\smallskip}
\# & Model & dof\tablefootmark{a} & $\chi^2_\mathrm{MAP}$\tablefootmark{b} & $\ln\mathcal{Z}$ & $\Delta\ln\mathcal{Z}$ & DS \\\noalign{\smallskip}
\hline\noalign{\smallskip}
\multicolumn{7}{p{\hsize}}{A. a flat-line model} \\\noalign{\smallskip}
{\it i} & Full model & 139 & 346.1 & 752.88 & 0 & -- \\\noalign{\smallskip}
\hline\noalign{\smallskip}
\multicolumn{7}{p{\hsize}}{B. a pure atmospheric model assuming equilibrium chemistry} \\\noalign{\smallskip}
{\it i} & Full model & 133 & 305.6 & 790.34 & 0 & -- \\\noalign{\smallskip}
\hline\noalign{\smallskip}
\multicolumn{7}{p{\hsize}}{C. a pure atmospheric model assuming free chemistry} \\\noalign{\smallskip}
{\it i} & Full model & 120 & 298.2 & 792.21 & 0 & -- \\
{\it ii} & No Na & 121 & 298.4 &792.03 & $-0.17$ & N/A \\ 
{\it iii} & No K & 121 & 297.9& 793.21 & 1.00 & $-2.0\sigma$ \\ 
{\it iv} & No TiO & 121 & 297.5 & 793.97 & 1.76 & $-2.4\sigma$ \\ 
{\it v} & No VO & 121 & 298.7 & 794.08 & 1.87 & $-2.5\sigma$ \\ 
{\it vi} & No $\rm H_2O$ & 121 & 304.1 & 789.40 & $-2.81$ & $2.9\sigma$\\ 
{\it vii} & No FeH & 121 &  298.3& 792.55 & 0.34 & N/A \\\noalign{\smallskip}
\hline\noalign{\smallskip}
\multicolumn{7}{p{\hsize}}{D. a pure stellar contamination model} \\\noalign{\smallskip}
{\it i} & Spots \& faculae & 130 & 300.6& 780.06 & 0 & -- \\
{\it ii} & Only spots & 134 & 297.9 & 784.51 & 4.45  & $-3.4\sigma$ \\
{\it iii} & Only faculae & 134 & 345.0 & 769.74 & $-10.32$ & $4.9\sigma$ \\\noalign{\smallskip}
\hline\noalign{\smallskip}
\multicolumn{7}{p{\hsize}}{E. a hybrid model combining an atmosphere with equilibrium chemistry and stellar contamination} \\\noalign{\smallskip}
{\it i} & Full model & 128 & 238.0 & 806.72 & 0 & -- \\\noalign{\smallskip}
\hline\noalign{\smallskip}
\multicolumn{7}{p{\hsize}}{F. a hybrid model combining an atmosphere with free chemistry and stellar contamination} \\\noalign{\smallskip}
{\it i} & Full model & 115 & 226.6 & 807.96 & 0 & -- \\
{\it ii} & No Na & 116 & 227.4 & 809.65 & 1.69 & $-2.4\sigma$ \\ 
{\it iii} & No K & 116 & 230.6 & 807.99 & 0.02 & N/A \\ 
{\it iv} & No TiO & 116 & 230.0 & 810.30 & 2.33 & $-2.7\sigma$ \\ 
{\it v} & No VO & 116 & 226.7 & 810.69 & 2.72 & $-2.8\sigma$ \\ 
{\it vi} & No $\rm H_2O$ & 116 & 240.6 & 802.05 & $-5.91$ & $3.9\sigma$ \\
{\it vii} & No FeH & 116 & 232.1 & 807.48 & $-0.49$ & N/A \\\noalign{\smallskip}
\hline
\end{tabular}}
\tablefoot{
\tablefoottext{a}{Degrees of freedom.} 
\tablefoottext{b}{$\chi^2$ for the Maximum a Posteriori (MAP) model. }
}
\end{table}

\begin{table*}
\renewcommand\arraystretch{1.5} 
\caption[]{Priors and derived posteriors for the parameters used in the spectral retrieval analyses.}
\begin{center}
\scalebox{0.85}{
\begin{tabular}{cccccccc}
\hline
Parameter & Prior  & \multicolumn{5}{c}{Posterior}\\
 &   & Model A.{\it i} & Model B.{\it i} & Model C.{\it i} & Model D.{\it ii} & Model E.{\it i} & Model F.{\it i} \\
\hline
$T_\mathrm{iso}$ (K) 
& ${\mathcal{U}}(500,1500)$ 
& -- 
& $702^{+63}_{-85}$ 
& $637^{+86}_{-61}$
& -- 
& $608^{+37}_{-37}$ 
& $706^{+77}_{-66}$ \\
$\log P_0$ ($\log {\rm bar}$) 
& $\mathcal{U}(-6,2)$ 
& -- 
& $-4.90^{+1.07}_{-0.64}$ 
& $-4.94^{+0.52}_{-0.56}$
& -- 
& $-3.98^{+0.57}_{-0.89}$ 
& $-5.43^{+0.51}_{-0.35}$ \\
$\log P_\mathrm{cloud}$ ($\log {\rm bar}$) 
& $\mathcal{U}(-6,2)$ 
& -- 
& $0.32^{+1.18}_{-2.12}$ 
& $-0.41^{+1.55}_{-2.09}$
& -- 
& $1.51^{+0.32}_{-0.45}$ 
& $-0.45^{+1.45}_{-2.38}$ \\
$\log A_\mathrm{RS}$ 
& $\mathcal{U}(-4,4)$ 
& -- 
& $1.42^{+1.49}_{-3.51}$ 
& $0.68^{+1.91}_{-2.21}$
& -- 
& $-2.27^{+0.46}_{-0.55}$ 
& $-0.66^{+1.95}_{-1.86}$\\
$\phi$ 
& $\mathcal{U}(0,1)$ 
& -- 
& $0.33^{+0.16}_{-0.18}$
& $0.16^{+0.23}_{-0.11}$
& -- 
& $0.85^{+0.10}_{-0.18}$ 
& $0.15^{+0.22}_{-0.11}$ \\
$R_\mathrm{p}$ ($R_\mathrm{Jup}$) \tablefootmark{a}
& $\mathcal{U}(-0.5,1.5)$ 
& $1.063^{+0.001}_{-0.001}$ 
& --
& -- 
& $1.051^{+0.009}_{-0.035}$ 
& --
& -- \\
$T_\mathrm{phot}$ (K)
& $\mathcal{N}(4987,41)$ 
& -- 
& --
& -- 
& $4991^{+42}_{-41}$ 
& $4991^{+38}_{-38}$ 
& $4987^{+33}_{-34}$ \\
$\Delta T_\mathrm{spot}^\mathrm{N1}$ (K)
& $\mathcal{U}(-2500,2500)$ 
& -- 
& --
& -- 
& $-1617^{+332}_{-225}$ 
& $-1009^{+118}_{-187}$ 
& $-947^{+108}_{-107}$ \\
$f_\mathrm{spot}^\mathrm{N1}$ 
& $\mathcal{U}(0,1)$ 
& -- 
& --
& -- 
& $0.17^{+0.03}_{-0.04}$ 
& $0.24^{+0.04}_{-0.04}$ 
& $0.26^{+0.03}_{-0.03}$ \\
$\Delta T_\mathrm{spot}^\mathrm{N2}$ (K)
& $\mathcal{U}(-2500,2500)$ 
& -- 
& --
& -- 
& $-2026^{+1604}_{-360}$ 
& $-250^{+159}_{-340}$ 
& $-136^{+78}_{-126}$ \\
$f_\mathrm{spot}^\mathrm{N2}$ 
& $\mathcal{U}(0,1)$ 
& -- 
& --
& -- 
& $0.04^{+0.01}_{-0.01}$ 
& $0.09^{+0.10}_{-0.05}$ 
& $0.13^{+0.10}_{-0.06}$ \\
$\mathrm{C/O}$ 
& $\mathcal{U}(0.1,1.6)$ 
& -- 
& $0.14^{+0.07}_{-0.03}$ 
& -- 
& -- 
& $0.13^{+0.06}_{-0.02}$ 
& -- \\
$\log Z$ ($Z_\odot$)
& $\mathcal{U}(-2,3)$ 
& -- 
& $0.74^{+0.41}_{-1.21}$ 
& -- 
& -- 
& $-0.99^{+0.64}_{-0.33}$ 
& -- \\
$\log X_\mathrm{Na}$ 
& $\mathcal{U}(-12,0)$  
& -- 
& -- 
& $-8.47^{+1.19}_{-1.91}$
& -- 
& -- 
& $-9.03^{+1.41}_{-1.78}$ \\
$\log X_\mathrm{K}$ 
& $\mathcal{U}(-12,0)$  
& -- 
& -- 
& $-10.04^{+1.76}_{-1.32}$
& -- 
& -- 
& $-6.05^{+0.59}_{-1.03}$ \\
$\log X_\mathrm{TiO}$ 
& $\mathcal{U}(-12,0)$  
& -- 
& -- 
& $-11.24^{+0.57}_{-0.46}$
& -- 
& -- 
& $-11.37^{+0.50}_{-0.39}$ \\
$\log X_\mathrm{VO}$ 
& $\mathcal{U}(-12,0)$ 
& -- 
& -- 
& $-10.93^{+0.81}_{-0.70}$
& -- 
& -- 
& $-10.98^{+0.78}_{-0.68}$ \\
$\log X_\mathrm{H_2O}$ 
& $\mathcal{U}(-12,0)$ 
& -- 
& -- 
& $-1.04^{+0.24}_{-0.34}$
& -- 
& -- 
& $-0.79^{+0.14}_{-0.17}$ \\
$\log X_\mathrm{CH_4}$ 
& $\mathcal{U}(-12,0)$ 
& -- 
& -- 
& $-7.68^{+2.72}_{-2.70}$
& -- 
& -- 
& $-7.25^{+2.91}_{-3.00}$ \\
$\log X_\mathrm{CO}$ 
& $\mathcal{U}(-12,0)$ 
& -- 
& -- 
& $-6.73^{+3.26}_{-3.31}$
& -- 
& -- 
& $-6.74^{+3.26}_{-3.10}$ \\
$\log X_\mathrm{CO_2}$ 
& $\mathcal{U}(-12,0)$ 
& -- 
& -- 
& $-6.67^{+3.39}_{-3.31}$ 
& -- 
& -- 
& $-6.83^{+3.28}_{-3.18}$ \\
$\log X_\mathrm{HCN}$ 
& $\mathcal{U}(-12,0)$ 
& -- 
& -- 
& $-7.19^{+2.91}_{-2.98}$
& -- 
& -- 
& $-7.04^{+2.97}_{-2.85}$ \\
$\log X_\mathrm{C_2H_2}$ 
& $\mathcal{U}(-12,0)$ 
& -- 
& -- 
& $-6.29^{+3.04}_{-3.51}$
& -- 
& -- 
& $-6.76^{+3.34}_{-3.34}$ \\
$\log X_\mathrm{PH_3}$ 
& $\mathcal{U}(-12,0)$ 
& -- 
& -- 
& $-6.56^{+3.31}_{-3.37}$
& -- 
& -- 
& $-6.63^{+3.29}_{-3.29}$ \\
$\log X_\mathrm{NH_3}$ 
& $\mathcal{U}(-12,0)$ 
& -- 
& -- 
& $-7.60^{+2.61}_{-2.79}$ 
& -- 
& -- 
& $-7.37^{+2.77}_{-2.74}$ \\
$\log X_\mathrm{H_2S}$ 
& $\mathcal{U}(-12,0)$ 
& -- 
& -- 
& $-6.48^{+3.38}_{-3.43}$
& -- 
& -- 
& $-6.53^{+3.29}_{-3.38}$ \\
$\log X_\mathrm{SiO}$ 
& $\mathcal{U}(-12,0)$ 
& -- 
& -- 
& $-6.63^{+3.24}_{-3.22}$
& -- 
& -- 
& $-6.55^{+3.31}_{-3.25}$ \\
$\log X_\mathrm{FeH}$ 
& $\mathcal{U}(-12,0)$ 
& -- 
& -- 
& $-7.36^{+1.81}_{-2.40}$
& -- 
& -- 
& $-5.59^{+0.89}_{-1.57}$ \\
$\Delta_\mathrm{offset}$ (ppm)
& $\mathcal{U}(-5000,5000)$ 
& $53^{+66}_{-59}$ 
& $75^{+63}_{-60}$ 
& $83^{+58}_{-60}$ 
& $3136^{+1415}_{-114}$ 
& $4054^{+622}_{-545}$ 
& $4360^{+447}_{-396}$ \\
\hline
\end{tabular}
}
\tablefoot{
\tablefoottext{a}{In the flat-line and pure stellar contamination models, the baseline transit depth was parameterized via the planetary radius, $R_\mathrm{p}$. For models incorporating atmospheric assumptions, the transit depth was directly fitted based on the atmospheric parameters, and $R_\mathrm{p}$ was omitted from the parameter set.}
}
\label{tab:retrieved_param}  
\end{center}
\end{table*}

\begin{figure*}
\centering
\includegraphics[width=0.95\hsize]{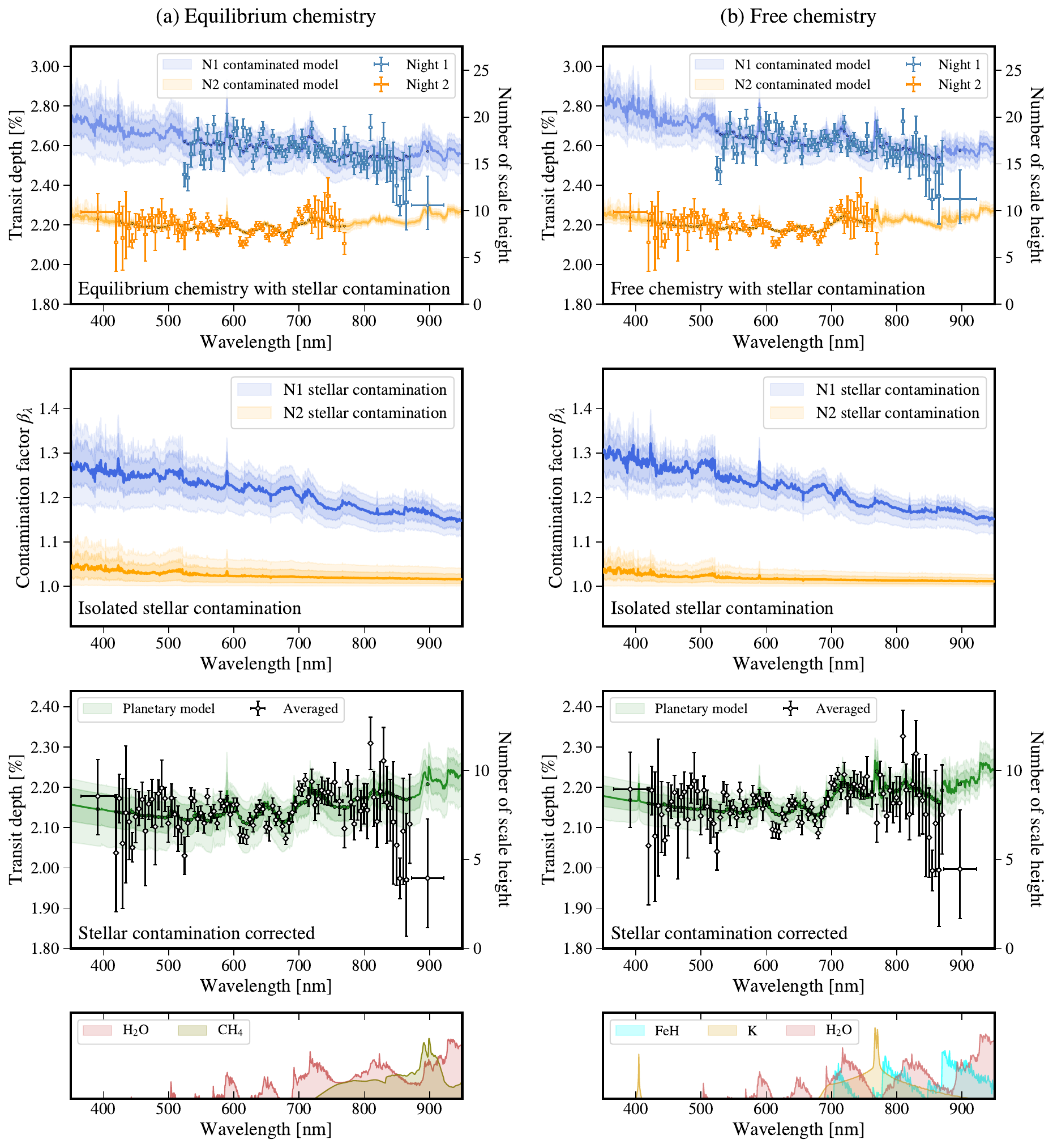}
\caption{Transmission spectra of NGTS-5b with the median, $1\sigma$, and $2\sigma$ confidence intervals of the retrieved hybrid model assuming equilibrium chemistry (Model E, left) and free chemistry (Model F, right). Top row: transmission spectra and retrieved models incorporating both atmospheric and stellar contamination effects, after correcting for the best-fit vertical offset ($\Delta_{\rm offset}$) derived from the joint retrieval. Second row: wavelength-dependent stellar contamination factors inferred from the retrieval. Third row: transmission spectra and corresponding models after correcting for stellar contamination. Bottom row: reference models illustrating the individual contributions of key species. }
\label{fig:retrieval_hybird}
\end{figure*}

\begin{figure}
\centering
\includegraphics[width=1.0\hsize]{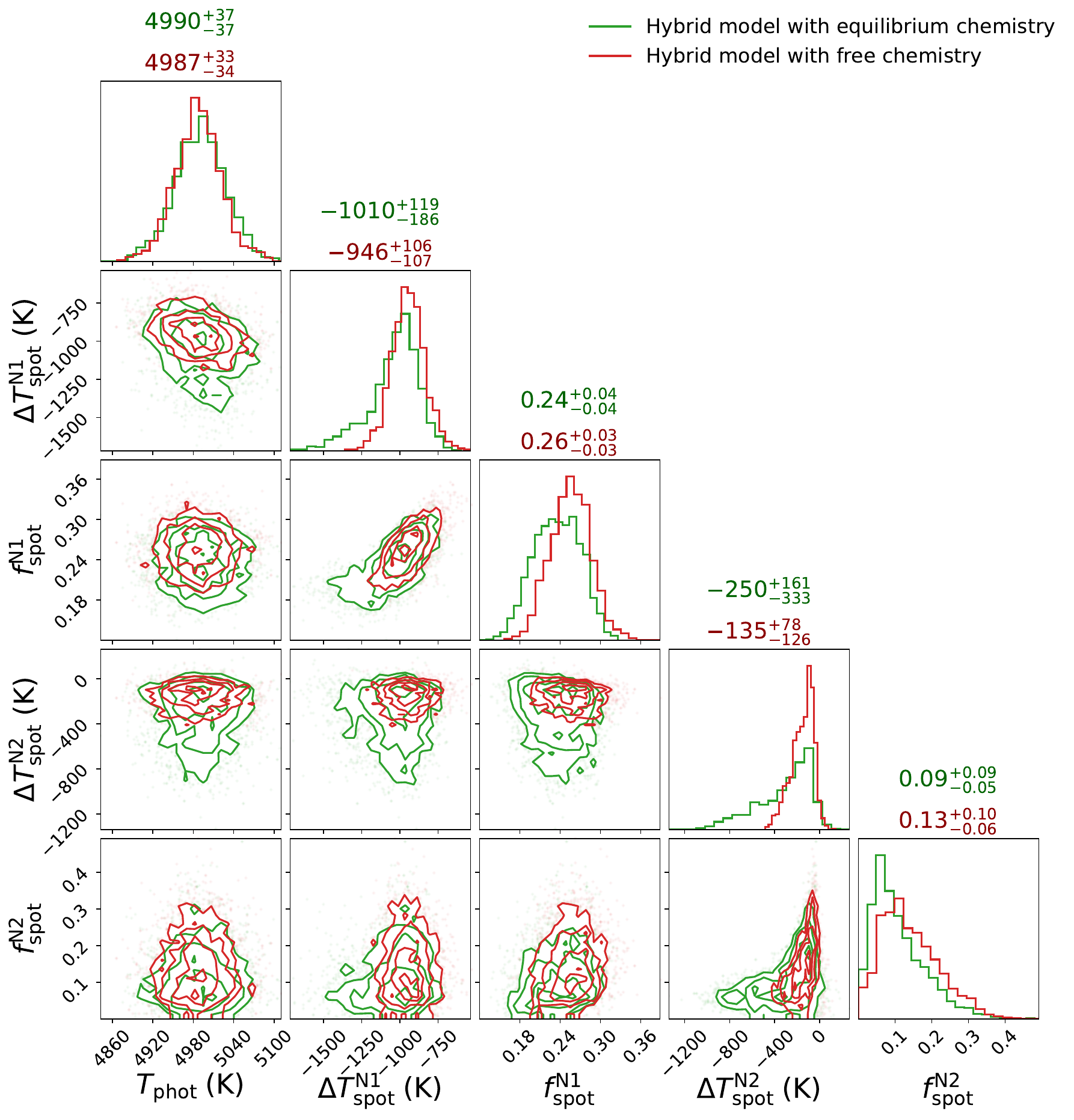}
\caption{Posterior distributions of the stellar contamination parameters for the retrieved hybrid model assuming equilibrium chemistry (Model E) and free chemistry (Model F). The diagonal panels show the marginalized distributions for each parameter. Detailed posterior values are listed in Table \ref{tab:retrieved_param}.}
\label{fig:corner_hybird_cmp}
\end{figure}

\begin{figure}
\centering
\includegraphics[width=1.0\hsize]{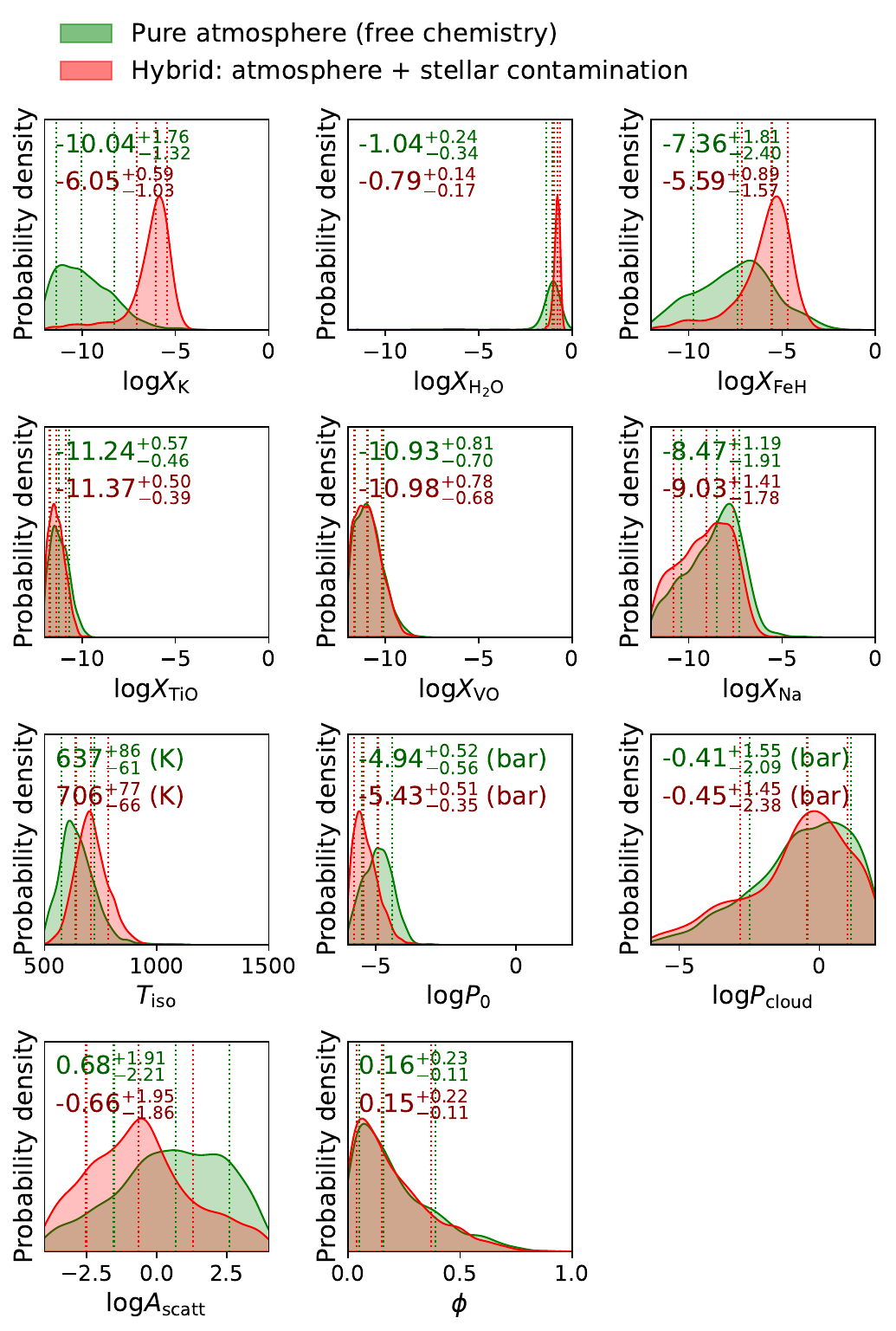}
\caption{Posterior distributions of the retrieved planetary atmosphere properties assuming free chemistry, comparing results with (red) and without (green) accounting for stellar contamination. Dashed lines mark the median values and the corresponding 1$\sigma$ credible intervals. Unconstrained parameters are not shown.}
\label{fig:post_sc_cmp}
\end{figure}

\subsection{Retrieval results}
\subsubsection{Evidence for stellar contamination}
Assuming an atmospheric scale height of $H/R_\star = (1.52 \pm 0.25) \times 10^{-3}$, the transmission spectrum from Night 1 (522--922~nm) spans $\sim 9$ scale heights, while that from Night 2 (366--772~nm) spans $\sim 5$. Using the retrieval framework described above, we investigated the contributions of planetary atmosphere and stellar contamination in the transmission spectra of NGTS-5b. Summary statistics and retrieved parameters for all models are listed in Tables \ref{tab:model_cmp} and \ref{tab:retrieved_param}. In Table \ref{tab:model_cmp}, detection significance (DS) is expressed in terms of the frequentist “sigma” significance level. Following \citet{Benneke2013}, the Bayes factor $\mathcal{B}_{10}$ can be related to the frequentist $p$-value $\rho$ via $\mathcal{B}_{10} \leq -\frac{1}{e\cdot\rho\cdot\ln\rho}$, where $e$ is the exponential of one and $\rho$ is the $p$-value. The $p$-value is then converted to the corresponding sigma significance $n_\sigma$ by $\rho = 1 - \operatorname{erf}\left({\frac{n_\sigma}{\sqrt{2}}}\right)$, where $\operatorname{erf}$ is the error function. This conversion is valid for $\rho < e^{-1}$ (i.e., $\lvert \Delta \ln \mathcal{Z} \rvert \geq 1.0$). For models with $\lvert \Delta \ln \mathcal{Z} \rvert < 1.0$, the corresponding significance is not reported and labeled as N/A. 

We first assessed pure atmospheric models. While both equilibrium chemistry (Model B) and free chemistry (Model C) improve the fit relative to a flat spectrum (Model A), neither can account for the discrepancy between the two nights in the common wavelength range. We therefore introduced epoch-dependent stellar contamination in the model assumptions. Bayesian model comparison indicates that hybrid models incorporating only unocculted spots are preferred. The model with spots alone (Model D.{\it ii}) is moderately favored over the model including both spots and faculae (Model D.{\it i}; $\Delta\ln\mathcal{Z} = 4.45$), while the latter is decisively favored over the faculae-only model (Model D.{\it iii}; $\Delta\ln\mathcal{Z} = 10.32$).

Compared to pure atmospheric models, hybrid models combining atmospheric features with unocculted spot contamination yield significantly improved fits and higher Bayesian evidences, with $\Delta\ln\mathcal{Z} = 16.38$ for the equilibrium chemistry case (Model E.{\it i}) and $\Delta\ln\mathcal{Z} = 15.75$ for free chemistry (Model F.{\it i}). The corresponding retrieved transmission spectra are shown in {Fig.~\ref{fig:retrieval_hybird}. In contrast, omitting the atmospheric component (Model D) results in a substantial decrease in Bayesian evidence ($\Delta\ln\mathcal{Z} > 22$), providing decisive evidence that both atmospheric signatures and stellar contamination are essential to explain the observed wavelength-dependent transit depth variations. Incorporating spot contamination also substantially reduces the night-to-night discrepancy, improving the $\chi^2$ from 103.03 to 73.18 after correcting for stellar contamination and applying a vertical offset (model F; with dof = 49). 

The retrievals suggest enhanced stellar activity during Night 1, characterized by a larger spot-to-photosphere temperature contrast ($-1196$ to $-839$~K) and higher spot coverage fraction (20–29\%) compared to Night 2, where the inferred contrast and coverage range from $-590$ to $-58$~K and 4–23\%, respectively. This behavior reflects the known degeneracy between $\Delta T_\mathrm{spot}$ and $f_\mathrm{spot}$, as illustrated in Fig. \ref{fig:corner_hybird_cmp}. These results demonstrate that stellar contamination significantly affects the transmission spectra of NGTS-5b, and can lead to biased inferences of planetary properties if not properly accounted for. 

Although no photometric signs of activity were found \citep{Eigmuller2019}, the Ca II H\&K lines indicate that NGTS-5 is active \citep[$\log R^\prime_\mathrm{HK}=-4.63$;][]{Vissapragada2022}. Given the lack of contemporaneous photometric monitoring with our OSIRIS observations, such epoch-to-epoch variations cannot be ruled out. Active regions on K-type stars can evolve on timescales of days to weeks, typically modulated by stellar rotation \citep{Zhang2020}. Similar epoch-to-epoch variations in stellar contamination have been reported for other warm Saturn- and sub-Saturn-mass planets orbiting K-type stars, such as HAT-P-12b \citep{Jiang2021}, HAT-P-18b \citep{Fournier-Tondreau2024,Perdelwitz2025}, and WASP-69b \citep{PetitditdelaRoche2024}. Discrepancies where photometric monitoring suggests lower activity levels than transmission spectroscopy are not uncommon in the literature \citep[e.g., for WASP-69b,][]{Murgas2020,2021AJ....162...91E,Ouyang2023,PetitditdelaRoche2024,Chakraborty2024}. However, while these epoch-to-epoch variations remain physically plausible, we caution that the activity inferred using the disk-integrated prescription of the Transit Light Source Effect \citep{Rackham2018,Rackham2019} might be overestimated due to degeneracies and model oversimplification \citep{2026arXiv260102621S}. 

\subsubsection{Constraints on the planetary atmosphere properties}
The atmospheric properties retrieved from the two hybrid models are broadly consistent with each other, with discrepancies within 1.5$\sigma$ for key parameters including the atmospheric temperature, reference pressure, cloud-top pressure, and haze enhancement factor.

The hybrid model assuming equilibrium chemistry yields well-constrained posteriors for the atmospheric temperature ($T_\mathrm{iso} = 608^{+37}_{-37}$~K), reference pressure ($\log P_0 = -3.98^{+0.57}_{-0.89}$~$\log {\rm bar}$), and haze enhancement factor ($\log A_\mathrm{RS} = -2.27^{+0.46}_{-0.55}$). The retrieved cloud-top pressure and cloud fraction posterior distributions are skewed towards the upper boundary of the prior ($\log P_\mathrm{cloud} > 0.81$~$\log {\rm bar}$, $\phi > 0.56$, at the $90\%$ lower limit), suggesting a relatively clear atmosphere with nearly uniform low-altitude cloud coverage. Additionally, the retrieval results in a low atmospheric metallicity of $0.10^{+0.34}_{-0.05}$ times solar and a C/O ratio skewed to the lower boundary of the prior ($\mathrm{C/O}<0.22$, $90\%$ upper limit).

The hybrid model adopting a free chemistry assumption produces atmospheric parameters consistent with the equilibrium chemistry case. Specifically, it retrieves $T_\mathrm{iso} = 706^{+77}_{-66}$~K, $\log P_0 < -4.75$~$\log {\rm bar}$ ($90\%$ upper limit), $\log P_\mathrm{cloud} = -0.45^{+1.45}_{-2.38}$~$\log {\rm bar}$, and $\log A_\mathrm{RS} = -0.66^{+1.95}_{-1.86}$. The cloud fraction posterior is skewed towards the lower prior bound ($\phi < 0.45$, $90\%$ upper limit), also indicating a relatively clear atmosphere. Among the 15 chemical species considered in the free chemistry model, only the mass fractions of K, $\rm H_2O$, and FeH are constrained, with retrieved values of  $\log X_\mathrm{K} = -6.05^{+0.59}_{-1.03}$, $\log X_\mathrm{H_2O} = -0.79^{+0.14}_{-0.17}$, and $\log X_\mathrm{FeH} = -5.59^{+0.89}_{-1.57}$, respectively.

To assess the individual contributions of these species, we performed a set of retrievals by sequentially removing each of the 15 species from the free chemistry model and comparing the resulting Bayesian evidence. As summarized in Table \ref{tab:model_cmp}, this analysis provides strong evidence for the presence of $\rm H_2O$ ($\Delta \ln \mathcal{Z} = 5.91$), while the evidence for K ($\Delta \ln \mathcal{Z} = -0.02$) and FeH ($\Delta \ln \mathcal{Z} = 0.49$) remains inconclusive.

Finally, to quantify the impact of stellar contamination on the atmospheric retrieval, we compared the posterior distributions of atmospheric parameters obtained from models with and without accounting for stellar contamination (Fig. \ref{fig:post_sc_cmp}). The inclusion of stellar contamination leads to more tightly constrained and systematically higher retrieved mass fractions of $\rm H_2O$, K, and FeH. This underscores the importance of properly accounting for stellar contamination. Neglecting such effects can result in the suppression of spectral features and biased atmospheric inferences.

\section{Discussion}
\label{Section:discussion}
Our retrieval analyses of the transmission spectra suggest that NGTS-5b possesses a relatively clear atmosphere dominated by water absorption, regardless of whether equilibrium or free chemistry is assumed. Given its equilibrium temperature of $952\pm 24$~K and mass of $0.229\pm 0.037$~$M_\mathrm{Jup}$ \citep{Eigmuller2019}, this result is consistent with NGTS-5b's proximity to the so-called ``clear sky corridor'' \citep{Ashtari2025} spanning 700--1700~K, where the 1.4~$\mu$m $\rm H_2O$ absorption band is typically stronger. 

Under the equilibrium chemistry assumption, our retrieval yields subsolar values for both atmospheric metallicity ($0.10^{+0.34}_{-0.05}\times$ solar) and C/O ($0.13^{+0.06}_{-0.02}$). These values may be biased due to the limited wavelength coverage of GTC/OSIRIS in the optical, where H$_2$O bands are relatively weaker than in the infrared, and where the absence of detectable carbon-bearing species limits the constraints on carbon abundance. Assuming the host star's metallicity of $0.12\pm 0.1$~dex, we find that NGTS-5b's mass and atmospheric metallicity are broadly consistent with the mass-metallicity relation established from exoplanet transmission spectra \citep{Welbanks2019, Sun2024}. The deviation from the predicted bulk metallicity could indicate a locally well-mixed interior status \citep{Thorngren2016,Mordasini2016,Hasegawa2024,Swain2024}. 

The free chemistry retrieval yields a mass fraction of $\log X_\mathrm{H_2O} = -0.79^{+0.14}_{-0.17}$. 
This corresponds to a log volume mixing ratio ($\log VMR$) of $-1.62^{+0.17}_{-0.19}$, which is comparable to values frequently retrieved in recent JWST studies, such as those for WASP-80b \citep[$-1.80^{+0.55}_{-0.94}$;][]{Bell2023}, WASP-94b \citep[$-1.59^{+0.35}_{-0.64}$;][]{2025MNRAS.540.2535A}, WASP-107b \citep[$-1.72^{+0.26}_{-0.25}$;][]{2024Natur.630..831S}, and HAT-P-12b \citep[$-2.09^{+0.58}_{-1.61}$;][]{2025A&A...703A.264C}. While HAT-P-12b \citep[0.21~$M_\mathrm{Jup}$, 963~K, K4 host;][]{Hartman2009} represents a very similar system in terms of both planetary mass and equilibrium temperature, another comaparable planet, HAT-P-18b \citep[0.20~$M_\mathrm{Jup}$, 852~K, K2 host;][]{Hartman2011}, exhibits a much lower H$_2$O abundance \citep[$-4.47^{+0.30}_{-0.28}$;][]{Fournier-Tondreau2024}, hinting at a potential atmospheric diversity among these warm Saturns. 

From the free chemistry retrieval, we derived a mean molecular weight of $\mu=2.7^{+0.19}_{-0.15}$. This indicates an H$_2$-dominated atmosphere and corresponds to a metallicity range of 3--50$\times$ solar under the assumption of chemical equilibrium, regardless of the adopted C/O ratio. 
This apparent inconsistency with the subsolar metallicity derived from the equilibrium chemistry retrieval likely arises from the limited spectral coverage and precision of current measurements, which restrict the ability to jointly constrain multiple key molecular species. Resolving this discrepancy will require additional observations to break the degeneracy between reference pressure, cloud property, metallicity, and molecular abundances.
As demonstrated by the benchmark planet WASP-39b, where retrieved values for H$_2$O ranged from $\log VMR\sim-$1.13 to $\log VMR\sim-$6.35 depending on JWST instrumental modes and atmospheric models, any interpretation of abundances should be approached with caution \citep{2024A&A...687A.110L}. 

We also note that the retrieved atmospheric temperatures of NGTS-5b under both equilibrium and free-chemistry assumptions are approximately 250–350~K lower than the corresponding equilibrium temperature. This discrepancy has been widely reported in exoplanet transmission spectroscopy. One possible explanation is that one-dimensional retrieval models can produce biased temperature estimates when applied to atmospheres with inhomogeneous terminator properties \citep{MacDonald2020}. Recent studies show that even planets with equilibrium temperatures as low as 770~K can exhibit substantial morning-to-evening limb asymmetry \citep[e.g.,][]{2024NatAs...8.1562M}.

Future infrared observations with JWST, capable of detecting the major carbon-, oxygen-, sulfur-, and nitrogen-bearing species, will be critical for confirming the presence of H$_2$O and for robustly constraining the atmospheric metallicity and elemental enrichment patterns. At the same time, the frequent presence of stellar contamination, as revealed in transmission spectroscopy, highlights the importance of complementary blue-optical observations, which have their strongest impact at ultraviolet and blue-optical wavelengths, where the contrast between active regions and quiescent photosphere is most pronounced \citep{Iyer2020,Fairman2024,Saba2025}. Together, these infrared and optical observations will enable precise measurements of elemental ratios (e.g., C/O) and metallicity (or [O/H]), ultimately providing insights into NGTS-5b’s formation and migration pathways \citep{Oberg2011,Madhusudhan2014a,Madhusudhan2017,Penzlin2024,Ohno2025}.

\section{Conclusions}
\label{Section:conclusions}
As a warm sub-Saturn located just above the upper boundary of the short-period Neptunian desert \citep{Mazeh2016}, potentially shaped by tidal disruption following high-eccentricity migration \citep{Matsakos2016,Owen2018}, NGTS-5b offers valuable opportunities to probe the formation and evolutionary history of close-in low-mass giant planets. In this work, we presented optical transmission spectra of NGTS-5b obtained from two transits observed with GTC/OSIRIS. Our analyses identified systematic discrepancies between the two observed spectra, which we attribute to varying levels of unocculted stellar activity across the two epochs.

By incorporating epoch-dependent stellar contamination alongside planetary atmospheric models in a joint retrieval framework, we found decisive Bayesian evidence ($\Delta \ln \mathcal{Z} > 15$) for the presence of stellar contamination, independent of atmospheric assumptions. The retrieved stellar activity parameters indicate enhanced activity during the first night, characterized by a larger spot-to-photosphere temperature contrast and higher spot coverage fraction relative to the second night. The discrepancies between the two transmission spectra were substantially reduced after correcting for these stellar contamination effects.

Atmospheric retrievals assuming both equilibrium and free chemistry consistently indicate a relatively clear atmosphere for NGTS-5b, with water vapor identified as the dominant opacity source. However, the retrieved atmospheric metallicities differ between assumptions, being subsolar under equilibrium chemistry and supersolar in the free chemistry case. We attribute this inconsistency to the limited wavelength coverage of GTC/OSIRIS, which includes only weak water bands and lacks diagnostic features from other key oxygen- and carbon-bearing species.

To robustly constrain the atmospheric composition and elemental ratios of NGTS-5b, future observations spanning a broader wavelength range, particularly in the infrared with JWST, complemented by blue-optical measurements with HST and large ground-based facilities, will be essential. Such comprehensive spectroscopic coverage will not only improve constraints on atmospheric metallicity and C/O ratio but also help mitigate the effects of stellar contamination, enabling a more complete understanding of the formation, migration, and evolutionary processes shaping this warm sub-Saturn.

\begin{acknowledgements}
We thank the anonymous referees for their constructive comments and suggestions. 
G.C. acknowledges the support by the National Natural Science Foundation of China (NSFC grant Nos. 42578016, 12122308, 42075122), Youth Innovation Promotion Association CAS (2021315), and the Minor Planet Foundation of the Purple Mountain Observatory. 
This work is based on observations made with the Gran Telescopio Canarias installed at the Spanish Observatorio del Roque de los Muchachos of the Instituto de Astrofísica de Canarias on the island of La Palma. 
\end{acknowledgements}

\bibliographystyle{aa.bst} 
\bibliography{reference.bib}

\begin{appendix}

\onecolumn
\section{Additional tables and figures}
%

\begin{longtable}{ccccccccccc}
\caption{Derived transmission spectra of NGTS-5b and the corresponding adopted limb-darkening coefficients for each spectroscopic channel.}\\
\renewcommand\arraystretch{1.5}
\label{tab:spectra} 
\endfirsthead
\caption{continued.}\\
\hline
$\lambda$ (nm) & \multicolumn{8}{c}{LDCs} & \multicolumn{2}{c}{$R_{\rm{p}}/R_\star$} \\
& \multicolumn{4}{c}{Night 1} & \multicolumn{4}{c}{Night 2} & Night 1 & Night 2 \\
&$u_1$ & $u_2$ & $u_3$ & $u_4$ & $u_1$ & $u_2$ & $u_3$ & $u_4$ & & \\
\hline
\endhead
\hline
\endfoot
\hline
$\lambda$ (nm) & \multicolumn{8}{c}{LDCs} & \multicolumn{2}{c}{$R_{\rm{p}}/R_\star$} \\
& \multicolumn{4}{c}{Night 1} & \multicolumn{4}{c}{Night 2} & Night 1 & Night 2 \\
&$u_1$ & $u_2$ & $u_3$ & $u_4$ & $u_1$ & $u_2$ & $u_3$ & $u_4$ & & \\
\hline
366--417 & -- & -- & -- & -- & $0.7924$ & $-1.4994$ & $2.4298$ & $-0.7457$ & -- & $0.1505^{+0.0030}_{-0.0032}$ \\
417--422 & -- & -- & -- & -- & $0.8480$ & $-1.5328$ & $2.3769$ & $-0.7193$ & -- & $0.1454^{+0.0050}_{-0.0050}$ \\
422--427 & -- & -- & -- & -- & $0.8329$ & $-1.5640$ & $2.5062$ & $-0.8079$ & -- & $0.1503^{+0.0020}_{-0.0020}$ \\
427--432 & -- & -- & -- & -- & $0.9558$ & $-1.4861$ & $2.0546$ & $-0.5715$ & -- & $0.1461^{+0.0055}_{-0.0055}$ \\
432--437 & -- & -- & -- & -- & $0.7963$ & $-1.3434$ & $2.2476$ & $-0.7394$ & -- & $0.1484^{+0.0056}_{-0.0059}$ \\
437--442 & -- & -- & -- & -- & $0.8139$ & $-1.3131$ & $2.1055$ & $-0.6507$ & -- & $0.1477^{+0.0028}_{-0.0032}$ \\
442--447 & -- & -- & -- & -- & $0.7077$ & $-1.1347$ & $2.1341$ & $-0.7551$ & -- & $0.1456^{+0.0013}_{-0.0012}$ \\
447--452 & -- & -- & -- & -- & $0.7014$ & $-1.1758$ & $2.2864$ & $-0.8528$ & -- & $0.1480^{+0.0023}_{-0.0022}$ \\
452--457 & -- & -- & -- & -- & $0.7051$ & $-1.1289$ & $2.1961$ & $-0.8196$ & -- & $0.1487^{+0.0014}_{-0.0014}$ \\
457--462 & -- & -- & -- & -- & $0.7345$ & $-1.1386$ & $2.1465$ & $-0.7924$ & -- & $0.1495^{+0.0014}_{-0.0015}$ \\
462--467 & -- & -- & -- & -- & $0.7113$ & $-1.0970$ & $2.1787$ & $-0.8416$ & -- & $0.1467^{+0.0030}_{-0.0046}$ \\
467--472 & -- & -- & -- & -- & $0.7247$ & $-1.1341$ & $2.1828$ & $-0.8257$ & -- & $0.1490^{+0.0011}_{-0.0013}$ \\
472--477 & -- & -- & -- & -- & $0.7457$ & $-1.2248$ & $2.2759$ & $-0.8504$ & -- & $0.1496^{+0.0016}_{-0.0023}$ \\
477--482 & -- & -- & -- & -- & $0.7432$ & $-1.1950$ & $2.2760$ & $-0.8776$ & -- & $0.1472^{+0.0029}_{-0.0032}$ \\
482--487 & -- & -- & -- & -- & $0.7803$ & $-1.1510$ & $2.1069$ & $-0.7989$ & -- & $0.1499^{+0.0011}_{-0.0013}$ \\
487--492 & -- & -- & -- & -- & $0.8026$ & $-1.2321$ & $2.1863$ & $-0.8204$ & -- & $0.1504^{+0.0023}_{-0.0032}$ \\
492--497 & -- & -- & -- & -- & $0.8614$ & $-1.4289$ & $2.3942$ & $-0.8873$ & -- & $0.1496^{+0.0009}_{-0.0011}$ \\
497--502 & -- & -- & -- & -- & $0.8718$ & $-1.4297$ & $2.3593$ & $-0.8696$ & -- & $0.1476^{+0.0014}_{-0.0016}$ \\
502--507 & -- & -- & -- & -- & $0.9297$ & $-1.6456$ & $2.6006$ & $-0.9480$ & -- & $0.1499^{+0.0012}_{-0.0016}$ \\
507--512 & -- & -- & -- & -- & $1.0000$ & $-1.8357$ & $2.7535$ & $-0.9805$ & -- & $0.1486^{+0.0014}_{-0.0019}$ \\
512--517 & -- & -- & -- & -- & $0.9853$ & $-1.6473$ & $2.4070$ & $-0.8259$ & -- & $0.1475^{+0.0018}_{-0.0020}$ \\
517--522 & $0.9869$ & $-1.5972$ & $2.3548$ & $-0.8348$ & $0.9808$ & $-1.5846$ & $2.3461$ & $-0.8306$ & -- & $0.1471^{+0.0012}_{-0.0011}$ \\
522--527 & $0.8570$ & $-1.2532$ & $2.1046$ & $-0.7903$ & $0.8608$ & $-1.2617$ & $2.1089$ & $-0.7902$ & $0.1431^{+0.0022}_{-0.0021}$ & $0.1468^{+0.0022}_{-0.0026}$ \\
527--532 & $0.8890$ & $-1.3243$ & $2.1926$ & $-0.8396$ & $0.8898$ & $-1.3195$ & $2.1757$ & $-0.8289$ & $0.1426^{+0.0024}_{-0.0023}$ & $0.1488^{+0.0013}_{-0.0012}$ \\
532--537 & $0.8689$ & $-1.2583$ & $2.1152$ & $-0.8122$ & $0.8763$ & $-1.2779$ & $2.1297$ & $-0.8137$ & $0.1467^{+0.0020}_{-0.0023}$ & $0.1501^{+0.0009}_{-0.0010}$ \\
537--542 & $0.9170$ & $-1.3598$ & $2.1566$ & $-0.8060$ & $0.9320$ & $-1.3868$ & $2.1727$ & $-0.8113$ & $0.1487^{+0.0021}_{-0.0022}$ & $0.1495^{+0.0009}_{-0.0010}$ \\
542--547 & $0.9011$ & $-1.3462$ & $2.1987$ & $-0.8414$ & $0.8962$ & $-1.3300$ & $2.1845$ & $-0.8388$ & $0.1509^{+0.0020}_{-0.0019}$ & $0.1470^{+0.0012}_{-0.0013}$ \\
547--552 & $0.9170$ & $-1.3523$ & $2.1650$ & $-0.8212$ & $0.9174$ & $-1.3572$ & $2.1774$ & $-0.8289$ & $0.1469^{+0.0018}_{-0.0020}$ & $0.1483^{+0.0006}_{-0.0007}$ \\
552--557 & $0.8823$ & $-1.2643$ & $2.1239$ & $-0.8328$ & $0.8934$ & $-1.2740$ & $2.0967$ & $-0.8096$ & $0.1457^{+0.0017}_{-0.0017}$ & $0.1477^{+0.0010}_{-0.0009}$ \\
557--562 & $0.9248$ & $-1.3360$ & $2.1448$ & $-0.8287$ & $0.9143$ & $-1.3193$ & $2.1406$ & $-0.8309$ & $0.1502^{+0.0015}_{-0.0018}$ & $0.1496^{+0.0008}_{-0.0008}$ \\
562--567 & $0.8948$ & $-1.2424$ & $2.0555$ & $-0.8056$ & $0.9027$ & $-1.2670$ & $2.0900$ & $-0.8223$ & $0.1458^{+0.0016}_{-0.0017}$ & $0.1483^{+0.0008}_{-0.0007}$ \\
567--572 & $0.9283$ & $-1.3165$ & $2.1224$ & $-0.8328$ & $0.9361$ & $-1.3268$ & $2.1147$ & $-0.8242$ & $0.1517^{+0.0015}_{-0.0016}$ & $0.1477^{+0.0006}_{-0.0006}$ \\
572--577 & $0.9062$ & $-1.2906$ & $2.1200$ & $-0.8342$ & $0.9036$ & $-1.2809$ & $2.1015$ & $-0.8240$ & $0.1489^{+0.0016}_{-0.0015}$ & $0.1469^{+0.0006}_{-0.0006}$ \\
577--582 & $0.9130$ & $-1.2872$ & $2.1078$ & $-0.8350$ & $0.9179$ & $-1.2914$ & $2.1073$ & $-0.8349$ & $0.1487^{+0.0019}_{-0.0019}$ & $0.1491^{+0.0008}_{-0.0009}$ \\
582--587 & $0.9311$ & $-1.3200$ & $2.1242$ & $-0.8390$ & $0.9608$ & $-1.3983$ & $2.2013$ & $-0.8676$ & $0.1460^{+0.0024}_{-0.0022}$ & $0.1487^{+0.0006}_{-0.0006}$ \\
587--592 & $0.9879$ & $-1.4890$ & $2.2773$ & $-0.8845$ & $0.9827$ & $-1.4758$ & $2.2643$ & $-0.8802$ & $0.1518^{+0.0018}_{-0.0022}$ & $0.1488^{+0.0007}_{-0.0008}$ \\
592--597 & $0.9405$ & $-1.3280$ & $2.1134$ & $-0.8338$ & $0.9478$ & $-1.3466$ & $2.1315$ & $-0.8407$ & $0.1477^{+0.0015}_{-0.0016}$ & $0.1483^{+0.0011}_{-0.0015}$ \\
597--602 & $0.9623$ & $-1.3642$ & $2.1209$ & $-0.8328$ & $0.9616$ & $-1.3691$ & $2.1378$ & $-0.8414$ & $0.1506^{+0.0023}_{-0.0024}$ & $0.1485^{+0.0006}_{-0.0006}$ \\
602--607 & $0.9878$ & $-1.4265$ & $2.1813$ & $-0.8549$ & $0.9797$ & $-1.4052$ & $2.1642$ & $-0.8511$ & $0.1511^{+0.0013}_{-0.0014}$ & $0.1482^{+0.0006}_{-0.0006}$ \\
607--612 & $0.9850$ & $-1.4026$ & $2.1437$ & $-0.8425$ & $0.9871$ & $-1.4135$ & $2.1620$ & $-0.8505$ & $0.1477^{+0.0015}_{-0.0014}$ & $0.1454^{+0.0009}_{-0.0008}$ \\
612--617 & $1.0224$ & $-1.4757$ & $2.1742$ & $-0.8464$ & $1.0255$ & $-1.4834$ & $2.1852$ & $-0.8515$ & $0.1512^{+0.0014}_{-0.0017}$ & $0.1450^{+0.0007}_{-0.0007}$ \\
617--622 & $0.9928$ & $-1.4035$ & $2.1208$ & $-0.8328$ & $1.0013$ & $-1.4280$ & $2.1415$ & $-0.8376$ & $0.1489^{+0.0020}_{-0.0018}$ & $0.1455^{+0.0006}_{-0.0007}$ \\
622--627 & $1.0528$ & $-1.5643$ & $2.2653$ & $-0.8769$ & $1.0292$ & $-1.4991$ & $2.2077$ & $-0.8631$ & $0.1500^{+0.0011}_{-0.0011}$ & $0.1469^{+0.0006}_{-0.0007}$ \\
627--632 & $1.0096$ & $-1.4551$ & $2.1635$ & $-0.8433$ & $1.0257$ & $-1.4860$ & $2.1784$ & $-0.8445$ & $0.1455^{+0.0013}_{-0.0014}$ & $0.1471^{+0.0006}_{-0.0006}$ \\
632--637 & $1.0417$ & $-1.5382$ & $2.2543$ & $-0.8841$ & $1.0423$ & $-1.5405$ & $2.2578$ & $-0.8854$ & $0.1473^{+0.0011}_{-0.0011}$ & $0.1483^{+0.0006}_{-0.0007}$ \\
637--642 & $1.0790$ & $-1.6312$ & $2.3314$ & $-0.9060$ & $1.0881$ & $-1.6497$ & $2.3420$ & $-0.9079$ & $0.1492^{+0.0012}_{-0.0012}$ & $0.1483^{+0.0006}_{-0.0006}$ \\
642--647 & $1.0868$ & $-1.6299$ & $2.3149$ & $-0.9036$ & $1.0877$ & $-1.6434$ & $2.3314$ & $-0.9075$ & $0.1510^{+0.0012}_{-0.0011}$ & $0.1477^{+0.0005}_{-0.0005}$ \\
647--652 & $1.0794$ & $-1.6280$ & $2.3220$ & $-0.9063$ & $1.0605$ & $-1.5651$ & $2.2421$ & $-0.8754$ & $0.1480^{+0.0012}_{-0.0014}$ & $0.1463^{+0.0005}_{-0.0005}$ \\
652--657 & $1.0671$ & $-1.4967$ & $2.1285$ & $-0.8437$ & $1.0699$ & $-1.5114$ & $2.1575$ & $-0.8582$ & $0.1452^{+0.0018}_{-0.0013}$ & $0.1467^{+0.0015}_{-0.0012}$ \\
657--662 & $1.0567$ & $-1.5229$ & $2.1738$ & $-0.8493$ & $1.0724$ & $-1.5824$ & $2.2668$ & $-0.8940$ & $0.1498^{+0.0011}_{-0.0011}$ & $0.1474^{+0.0008}_{-0.0009}$ \\
662--667 & $1.0849$ & $-1.5994$ & $2.2624$ & $-0.8873$ & $1.0874$ & $-1.6112$ & $2.2776$ & $-0.8930$ & $0.1472^{+0.0011}_{-0.0011}$ & $0.1470^{+0.0006}_{-0.0006}$ \\
667--672 & $1.1037$ & $-1.6573$ & $2.3147$ & $-0.9034$ & $1.1075$ & $-1.6687$ & $2.3272$ & $-0.9076$ & $0.1481^{+0.0016}_{-0.0014}$ & $0.1477^{+0.0006}_{-0.0005}$ \\
672--677 & $1.0933$ & $-1.6087$ & $2.2579$ & $-0.8872$ & $1.0905$ & $-1.5892$ & $2.2181$ & $-0.8675$ & $0.1473^{+0.0010}_{-0.0011}$ & $0.1465^{+0.0005}_{-0.0005}$ \\
677--682 & $1.1171$ & $-1.6813$ & $2.3257$ & $-0.9082$ & $1.1099$ & $-1.6538$ & $2.2846$ & $-0.8901$ & $0.1463^{+0.0011}_{-0.0010}$ & $0.1454^{+0.0005}_{-0.0005}$ \\
682--687 & $1.1272$ & $-1.6984$ & $2.3385$ & $-0.9154$ & $1.1337$ & $-1.7146$ & $2.3508$ & $-0.9185$ & $0.1484^{+0.0013}_{-0.0013}$ & $0.1463^{+0.0012}_{-0.0011}$ \\
687--692 & $1.1226$ & $-1.7008$ & $2.3492$ & $-0.9210$ & $1.1254$ & $-1.7048$ & $2.3512$ & $-0.9215$ & $0.1488^{+0.0011}_{-0.0011}$ & $0.1475^{+0.0013}_{-0.0011}$ \\
692--697 & $1.1494$ & $-1.7570$ & $2.3863$ & $-0.9306$ & $1.1408$ & $-1.7315$ & $2.3628$ & $-0.9238$ & $0.1479^{+0.0012}_{-0.0010}$ & $0.1487^{+0.0008}_{-0.0009}$ \\
697--702 & $1.1501$ & $-1.7656$ & $2.3947$ & $-0.9335$ & $1.1422$ & $-1.7328$ & $2.3556$ & $-0.9201$ & $0.1473^{+0.0011}_{-0.0012}$ & $0.1506^{+0.0007}_{-0.0007}$ \\
702--707 & $1.1523$ & $-1.7455$ & $2.3478$ & $-0.9155$ & $1.1642$ & $-1.7770$ & $2.3766$ & $-0.9241$ & $0.1494^{+0.0015}_{-0.0016}$ & $0.1496^{+0.0006}_{-0.0006}$ \\
707--712 & $1.1841$ & $-1.8431$ & $2.4582$ & $-0.9562$ & $1.1852$ & $-1.8446$ & $2.4568$ & $-0.9547$ & $0.1487^{+0.0011}_{-0.0011}$ & $0.1510^{+0.0006}_{-0.0006}$ \\
712--717 & $1.1981$ & $-1.8749$ & $2.4831$ & $-0.9655$ & $1.1736$ & $-1.8019$ & $2.3985$ & $-0.9336$ & $0.1470^{+0.0010}_{-0.0011}$ & $0.1516^{+0.0011}_{-0.0011}$ \\
717--722 & $1.2088$ & $-1.8902$ & $2.4748$ & $-0.9569$ & $1.1990$ & $-1.8508$ & $2.4164$ & $-0.9324$ & $0.1510^{+0.0012}_{-0.0011}$ & $0.1489^{+0.0022}_{-0.0020}$ \\
722--727 & $1.1935$ & $-1.8693$ & $2.4819$ & $-0.9685$ & $1.1905$ & $-1.8604$ & $2.4724$ & $-0.9652$ & $0.1467^{+0.0016}_{-0.0014}$ & $0.1505^{+0.0027}_{-0.0025}$ \\
727--732 & $1.2095$ & $-1.8828$ & $2.4653$ & $-0.9594$ & $1.2081$ & $-1.8862$ & $2.4704$ & $-0.9595$ & $0.1474^{+0.0014}_{-0.0013}$ & $0.1493^{+0.0032}_{-0.0034}$ \\
732--737 & $1.2079$ & $-1.8976$ & $2.4778$ & $-0.9588$ & $1.2129$ & $-1.8967$ & $2.4698$ & $-0.9576$ & $0.1481^{+0.0012}_{-0.0012}$ & $0.1490^{+0.0032}_{-0.0030}$ \\
737--742 & $1.2333$ & $-1.9452$ & $2.5167$ & $-0.9755$ & $1.2266$ & $-1.9174$ & $2.4765$ & $-0.9598$ & $0.1471^{+0.0011}_{-0.0010}$ & $0.1510^{+0.0028}_{-0.0028}$ \\
742--747 & $1.2689$ & $-2.0365$ & $2.5956$ & $-1.0015$ & $1.2351$ & $-1.9573$ & $2.5324$ & $-0.9830$ & $0.1465^{+0.0013}_{-0.0012}$ & $0.1532^{+0.0030}_{-0.0029}$ \\
747--752 & $1.2111$ & $-1.8931$ & $2.4732$ & $-0.9645$ & $1.2287$ & $-1.9333$ & $2.5043$ & $-0.9745$ & $0.1467^{+0.0011}_{-0.0011}$ & $0.1497^{+0.0018}_{-0.0018}$ \\
752--757 & $1.2274$ & $-1.9342$ & $2.5080$ & $-0.9771$ & $1.2223$ & $-1.9129$ & $2.4821$ & $-0.9676$ & $0.1487^{+0.0023}_{-0.0022}$ & $0.1494^{+0.0027}_{-0.0022}$ \\
757--767 & $1.2444$ & $-1.9595$ & $2.5024$ & $-0.9706$ & $1.2323$ & $-1.9351$ & $2.4905$ & $-0.9692$ & $0.1462^{+0.0015}_{-0.0015}$ & $0.1491^{+0.0019}_{-0.0019}$ \\
767--772 & $1.2342$ & $-1.9433$ & $2.5049$ & $-0.9776$ & $1.2332$ & $-1.9400$ & $2.5004$ & $-0.9758$ & $0.1485^{+0.0030}_{-0.0035}$ & $0.1451^{+0.0018}_{-0.0019}$ \\
772--777 & $1.2341$ & $-1.9302$ & $2.4735$ & $-0.9645$ & $1.2268$ & $-1.9096$ & $2.4488$ & $-0.9550$ & $0.1489^{+0.0012}_{-0.0011}$ & -- \\
777--782 & $1.2565$ & $-1.9870$ & $2.5285$ & $-0.9850$ & $1.2492$ & $-1.9737$ & $2.5205$ & $-0.9821$ & $0.1469^{+0.0015}_{-0.0015}$ & -- \\
782--787 & $1.2573$ & $-1.9861$ & $2.5228$ & $-0.9824$ & $1.2604$ & $-1.9990$ & $2.5409$ & $-0.9897$ & $0.1445^{+0.0015}_{-0.0014}$ & -- \\
787--792 & $1.3001$ & $-2.1168$ & $2.6573$ & $-1.0284$ & -- & -- & -- & -- & $0.1467^{+0.0020}_{-0.0018}$ & -- \\
792--797 & $1.2796$ & $-2.0492$ & $2.5683$ & $-0.9940$ & -- & -- & -- & -- & $0.1452^{+0.0017}_{-0.0017}$ & -- \\
797--802 & $1.3154$ & $-2.1622$ & $2.6959$ & $-1.0419$ & -- & -- & -- & -- & $0.1456^{+0.0018}_{-0.0016}$ & -- \\
802--807 & $1.3078$ & $-2.1438$ & $2.6773$ & $-1.0349$ & -- & -- & -- & -- & $0.1449^{+0.0023}_{-0.0025}$ & -- \\
807--812 & $1.3447$ & $-2.2474$ & $2.7909$ & $-1.0795$ & -- & -- & -- & -- & $0.1512^{+0.0021}_{-0.0022}$ & -- \\
812--817 & $1.2906$ & $-2.0873$ & $2.6183$ & $-1.0189$ & -- & -- & -- & -- & $0.1459^{+0.0022}_{-0.0025}$ & -- \\
817--822 & $1.3122$ & $-2.1401$ & $2.6378$ & $-1.0161$ & -- & -- & -- & -- & $0.1436^{+0.0020}_{-0.0022}$ & -- \\
822--827 & $1.3367$ & $-2.2010$ & $2.7113$ & $-1.0494$ & -- & -- & -- & -- & $0.1463^{+0.0025}_{-0.0026}$ & -- \\
827--832 & $1.3109$ & $-2.1361$ & $2.6532$ & $-1.0311$ & -- & -- & -- & -- & $0.1493^{+0.0028}_{-0.0035}$ & -- \\
832--837 & $1.3822$ & $-2.3301$ & $2.8311$ & $-1.0878$ & -- & -- & -- & -- & $0.1453^{+0.0031}_{-0.0035}$ & -- \\
837--842 & $1.3016$ & $-2.0985$ & $2.5928$ & $-1.0075$ & -- & -- & -- & -- & $0.1449^{+0.0024}_{-0.0029}$ & -- \\
842--847 & $1.3402$ & $-2.2126$ & $2.7200$ & $-1.0549$ & -- & -- & -- & -- & $0.1436^{+0.0052}_{-0.0054}$ & -- \\
847--852 & $1.3656$ & $-2.2693$ & $2.7558$ & $-1.0645$ & -- & -- & -- & -- & $0.1411^{+0.0044}_{-0.0035}$ & -- \\
852--857 & $1.4049$ & $-2.3602$ & $2.8063$ & $-1.0754$ & -- & -- & -- & -- & $0.1376^{+0.0019}_{-0.0018}$ & -- \\
857--862 & $1.3378$ & $-2.2049$ & $2.7077$ & $-1.0526$ & -- & -- & -- & -- & $0.1425^{+0.0046}_{-0.0046}$ & -- \\
862--867 & $1.3554$ & $-2.2197$ & $2.6704$ & $-1.0315$ & -- & -- & -- & -- & $0.1382^{+0.0060}_{-0.0051}$ & -- \\
867--872 & $1.3914$ & $-2.3396$ & $2.8078$ & $-1.0799$ & -- & -- & -- & -- & $0.1438^{+0.0043}_{-0.0034}$ & -- \\
872--922 & $1.3697$ & $-2.2887$ & $2.7676$ & $-1.0737$ & -- & -- & -- & -- & $0.1376^{+0.0053}_{-0.0045}$ & -- \\
\hline
\end{longtable}

\FloatBarrier 
\twocolumn

\FloatBarrier 
\clearpage

\end{appendix}
\end{document}